\documentclass[usenatbib,onecolumn]{mnras}

\usepackage{graphicx}
\usepackage{wrapfig}
\usepackage{natbib}

\newcommand{\be}{\begin{equation}}
\newcommand{\ee}{\end{equation}}
\newcommand{\nn}{\mbox{} \nonumber \\ \mbox{} }
\newcommand{\ba}{\begin{eqnarray}}
\newcommand{\ea}{\end{eqnarray}}

\newcommand{\E}{{\bf E}}
\newcommand{\B}{{\bf B}}

\renewcommand{\v}{{\bf v}}

\renewcommand{\div}{{\rm \,div\,}}

\newcommand\eg{{\it{e.g.\ }}}

\newcommand{\Bf}{{magnetic field}}

\newcommand{\Ef}{{electric  field}}

\newcommand{\NS}{neutron star}
\newcommand{\NSs}{{neutron stars}}
\newcommand{\EM}{electromagnetic}

\newcommand{\ms}{magnetosphere}
\newcommand{\mss}{magnetospheres}

\newcommand{\LC}{light cylinder}

\title{Rotating neutron stars  without light cylinders}

\author[Maxim Lyutikov, Praveen Sharma ]{Maxim Lyutikov, Praveen Sharma\\
Department of Physics  and Astronomy, Purdue University,   525 Northwestern Avenue, West Lafayette, IN47907-2036, USA; 
\\
lyutikov@purdue.edu}

\begin{document}

\maketitle

\begin{abstract}
We find a class of  twisted and differentially  rotating   \NS\   \mss\ that do not have a \LC,   generate no wind and thus do not spin-down. The \ms\ is composed of embedded differentially rotating  flux surfaces, with the angular velocity decreasing as  $\Omega \propto 1/r$ (equivalently, becoming smaller at the foot-points closer to the axis of rotation). 
For each given North-South self-similar   twist profile   there is a set of self-similar   angular velocity profiles (limited from above) with  a ``smooth'',  dipolar-like  \Bf\ structure extending to infinity. For  spin parameters  larger than some critical value,  the \LC\ appears, \ms\ opens up,  and the  wind is generated.  
 \end {abstract}

\section{Introduction}

Magnetospheres of pulsars are highly magnetized and relativistically  rotating \citep{GJ}.  Though the initial set-up is very simple - rotating magnetized sphere  with dipolar \Bf\ -  the nonlinear effects of charges and currents on the open field lines modify the field  and make the problem highly non-linear \citep{1988Ap&SS.146..205B,1999ApJ...511..351C,2005PhRvL..94b1101G,2006ApJ...648L..51S}.

 Another complication comes from possible non-dipolar, current-carrying  magnetospheric fields in the closed zone \citep{tlk,2013ApJ...774...92P,2013arXiv1306.2264L}. Current-loaded \mss, \eg\ slightly twisted dipole (so that the northern foot-point  is at different azimuthal angle $\phi$ than the  southern foot-point) are not the same as vacuum multipoles (though they can be expanded as a sum of). Qualitatively, a given twist of a field line concentrates at the point of lowest guiding field. Hence at  the furthest point (along  closed field lines)  in a (nearly) dipolar \ms.   This is especially important for magnetar phenomena \citep{tlk,2013ApJ...774...92P,2013arXiv1306.2264L}. 

Yet another complication comes from possible differential rotation of the foot points. This is induced by the electron Hall dynamics in the  \NS\ crust \citep{RG,2014PhPl...21e2110W,2015MNRAS.453L..93G}. In a ``Solar flare''  model of magnetar activity \cite{2006MNRAS.367.1594L,2015MNRAS.447.1407L} slow, plastic  twisting of footprints leads to kink instabilities, generation of flares and Coronal Mass ejections. 

Overall the system   becomes very  complicated/nonlinear. This puts special emphasis on  possible (semi)-analytical solutions, that  solve  the complicated mathematics problem, and yet can be easily  traced to the first principles. In this paper we discuss a class of (quasi)-analytical solutions of pulsar \mss\ with a highly unusual properties: rotating and twisted  pulsar \mss\  without {\LC}s.

\section{Self-similar rotating and twisted magnetospheres}

The workhorse of  analytical investigation of pulsar \mss\  is the relativistic generalization of the Grad-Shafranov equation \citep{1967PhFl...10..137G,Shafranov,1973ApJ...182..951S,1973ApJ...180L.133M,BeskinBook}  which  expresses the  force balance in an  axially symmetric \ms. 
The conditions of zero-divergence of the \Bf, the ideal condition and axial symmetry allow the force balance to be expressed as a single scalar equation for the shape of the magnetic flux function $\Psi (r,\theta)$. There is a mathematical problem, though: 
the source terms (the poloidal current and in the case of differentially  rotating \ms\ the spin)  enter as  terms  that are functions of the solution itself, not as independent quantities. This leads to mathematical complication: the equation to be solved and the solution need to be found self-consistently. 

Few analytical roads remain. First, one can prescribe $I(\Psi)$ and solve for $\Psi$. For example linear relations  $I = \alpha \Psi$ lead to simple, linear  and  hence highly useful analytical solutions, spheromaks in spherical geometry. Alternatively, \cite{1994MNRAS.267..146L} suggested  a self-similar approach, where the structure has a power-law scaling with  the  spherical radius.   \citep[They were adopted to cylindrical coordinates by][]{2020JPlPh..86b9010L}.
Twisted configurations  of  \cite{1994MNRAS.267..146L} are especially useful for astrophysical magnetars \citep{tlk}. 

In what follows we generalize the non-linear self-similar solutions of twisted \mss\  of \cite{1994MNRAS.267..146L} to twisted and  differentially rotating  configurations. 

\subsection{Self-similar  twisted and rotating  \mss}
\label{Self-similar}

For a stationary axially symmetric configuration the Grad-Shafranov prescription involves field parametrization
\ba && 
\B =(\nabla  \Psi ) \times (\nabla  \phi ) + I(\Psi )  (\nabla  \phi )
\nn &&
\E = -\nabla \Phi(\Psi ) 
\nn &&
\Omega = - \Phi'.
\ea
The force-free force balance 
\be
\E \div \E + (\nabla \times \B) \times \B=0
\ee
gives  then the Grad-Shafranov equation \citep{1967PhFl...10..137G,Shafranov,1973ApJ...182..951S,BeskinBook}
\be
(1 - r^2 \sin^2 \theta  \Omega^2 ) \Delta \Psi  - 2 \frac{ \left(  r \partial_r \Psi  + \cot \theta \partial_\theta \Psi  \right) }{ r^2} + I I'- 
(\nabla \Psi)^2 r^2 \sin^2 \theta  \Omega \Omega' =0,
\label{GS} 
\ee
(factors of $4 \pi$ omitted). ($\Omega$ is not the rate of shearing of foot-points, it's an angular velocity of rotation of a give flux surface.) 

Let's look for self-similar solutions to (\ref{GS})  in a form of
\ba &&
\Psi = r^{-p} F(\theta)
\nn && 
I  \propto \Psi^{\alpha}
\nn &&
\Omega= C_2 \Psi^{\beta}
\ea
By dimensional analysis
\ba &&
\alpha = 1+1/p, 
\nn &&
\beta = 1/p
\ea
so that the radial scaling of all the components of the \Bf\ is $\propto  r^{-2-p}$:
The electric part has the same power-law scaling as the magnetic.  \citep[We also comment that for $p=0$ one  recovers Michel's solution $\Psi = 1-\cos\theta$, but   with arbitrary $\Omega(\theta)$, $I = (1/2)  \sin^2 \theta \Omega(\theta)$][]{2011PhRvD..83l4035L}.

The angular part of the  Grad-Shafranov equation (\ref{GS})   then obeys
%Lynden-Bell-Rotating
\ba &&
 \left(  p(1+p) F +C_1 F^{1+2/p} + (1-\mu^2) F^{\prime\prime} \right)  - 
 \nn &&
2 C_2 ^2  \frac{ (1+p)^2}{p^3} (1-\mu^2) F^{2/p-1} 
\left( p^3 F^2 + (1-\mu^2)  F^{\prime ,2} + p F \left(  (1-\mu^2)  F^{\prime\prime}  -2 \mu F'\right) \right)  =0
\label{main}
\ea
where 
$\mu =\cos\theta$. 
The first part in parenthesis is the \cite{1994MNRAS.267..146L}  term. The last line is the contribution from self-similar rotations of the \ms.

For given $F$ the fields are given by
\ba &&
\B = \left\{ {F'},  p {F},\frac{1}{2} \sqrt{\frac{C_1 p}{p+1}} { F^{1+1/p} } \right\}  r^{-2-p} {\sin\theta} ^{-1}, 
\nn && 
\E=  \left\{F, -\frac{ F'} {p},0 \right\} C_2 (1+p)  r^{-2-p} F^{1/p}
\label{EB}
\ea

Solutions to Eq.  (\ref{main}) are of the eigenvalue problem:  there are three conditions:
\ba && 
F' (0) =0, \, \mbox{zero radial \Bf\ at } \theta =\pi/2
\nn && 
F(1) =0, \, \mbox{zero \,  } B_\theta  \, {\mbox at } \, \theta =0
\nn &&
F' (1) =-2,  \, \mbox{ radial \Bf\ at the pole, normalized to the dipole field}
\label{cond}
\ea
The three boundary conditions (\ref{cond}) impose an eigenvalue constraint on the triad of radial index $p$, twist $C_1$ and rotation  $C_2$: given the two, the eigenvalue constraint fixes the third.

\begin{figure}
\includegraphics[width=0.99\textwidth]{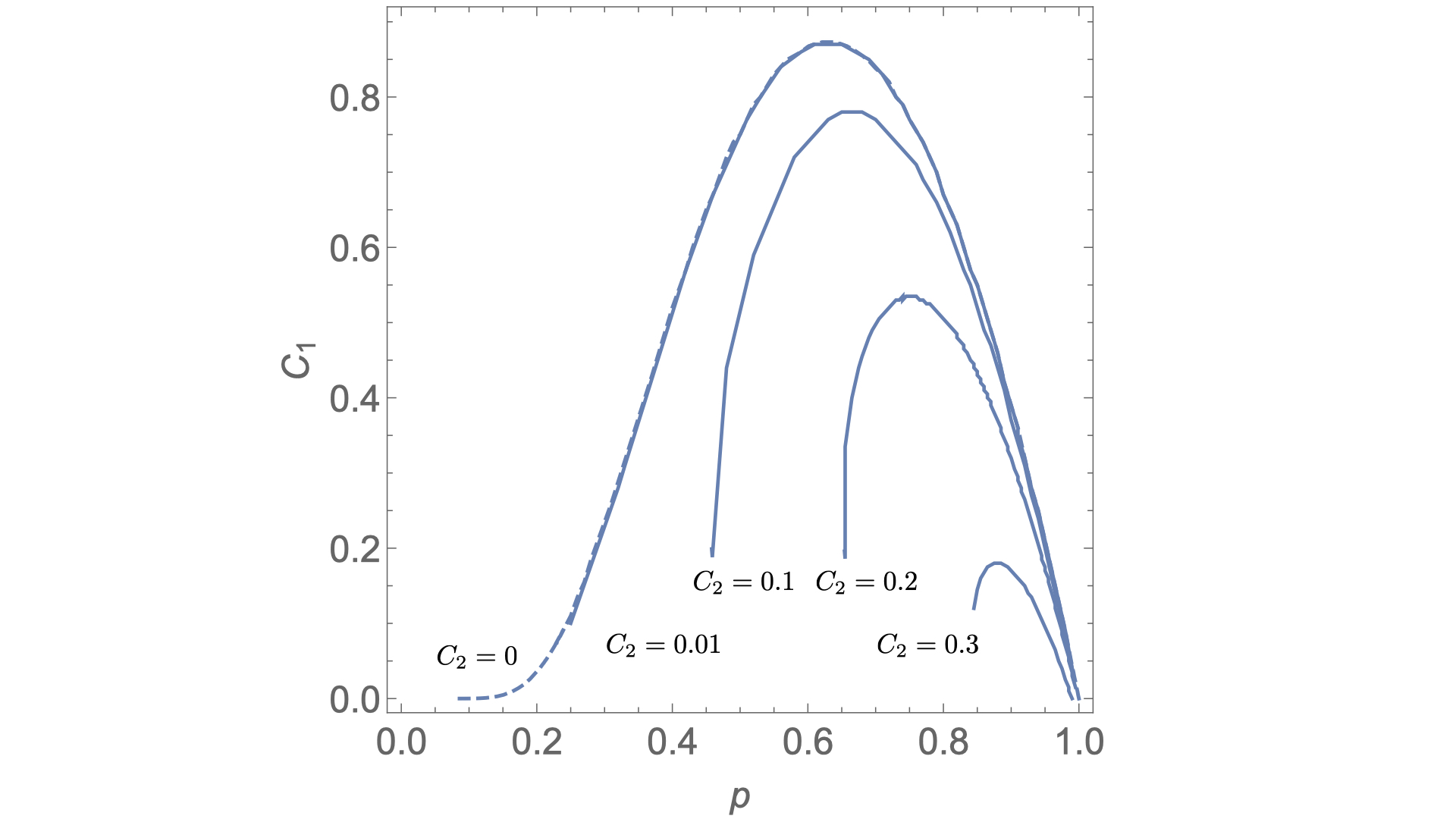}
\caption{Radial power-law index $p$ as function of the twist parameter $C_1$ for different spins parameters. Dashed line, non-rotating is   $C_2 = 0$. Other lines correspond to $C_2= 0.01,\,     0.1, \, 0.2,  \, 0.3$ (top to bottom). Note that for cases when there are no self-similar solutions (\eg\ for $p$ smaller than some critical value for each $C_2 \neq 0$) one may still  a have no-\LC\ \ms, but its structure is not self-similar.}
\label{NoLC2} 
  \end {figure}

\begin{figure}
\includegraphics[width=0.99\textwidth]{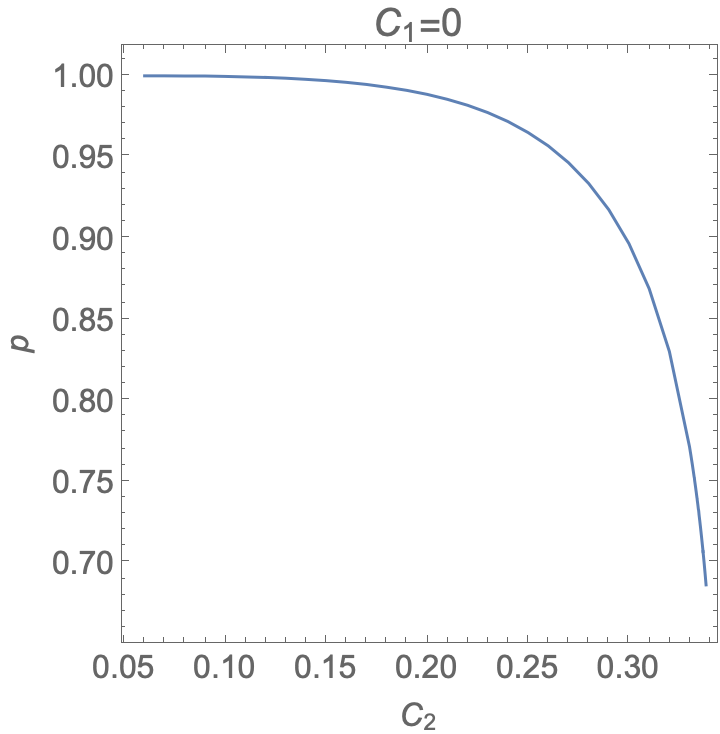}
\caption{ Radial power-law index $p$  as function of the rotation parameter $C_2$ for untwisted configuration  $C_1=0$.}
\label{c1-00} 
  \end {figure}

\begin{figure}
\includegraphics[width=0.33\textwidth]{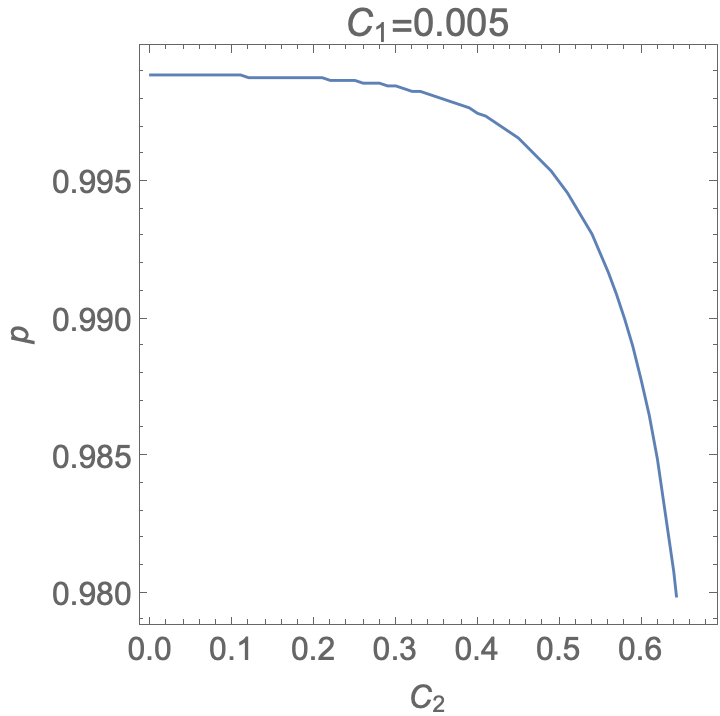}
\includegraphics[width=0.33\textwidth]{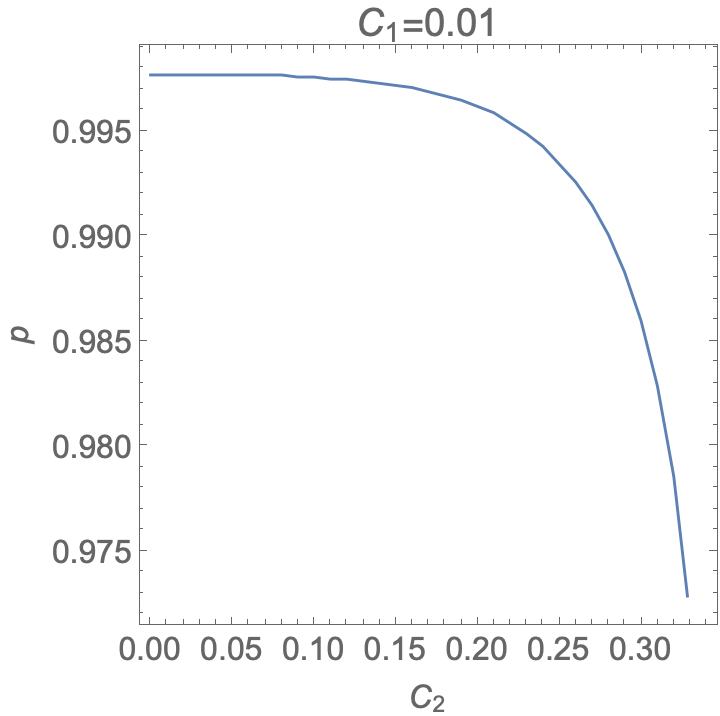}
\includegraphics[width=0.33\textwidth]{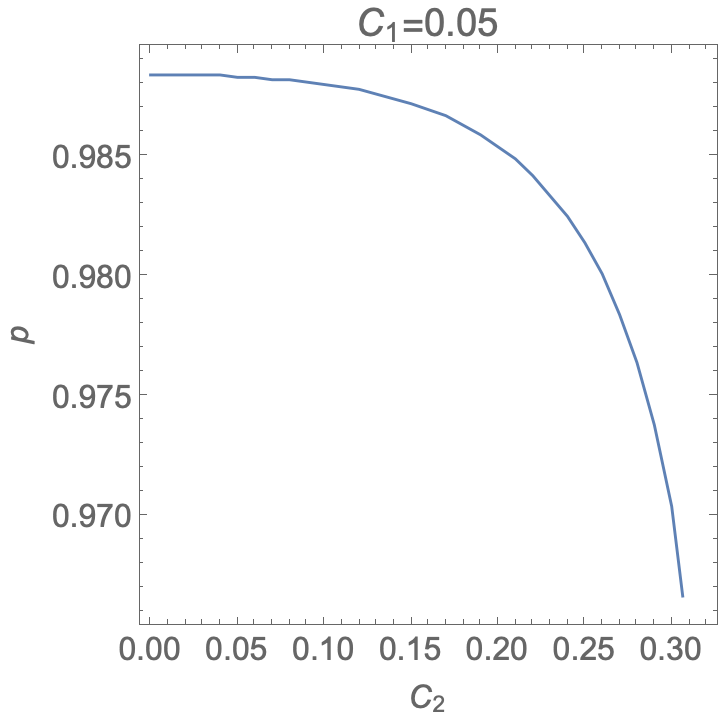}\\
\includegraphics[width=0.33\textwidth]{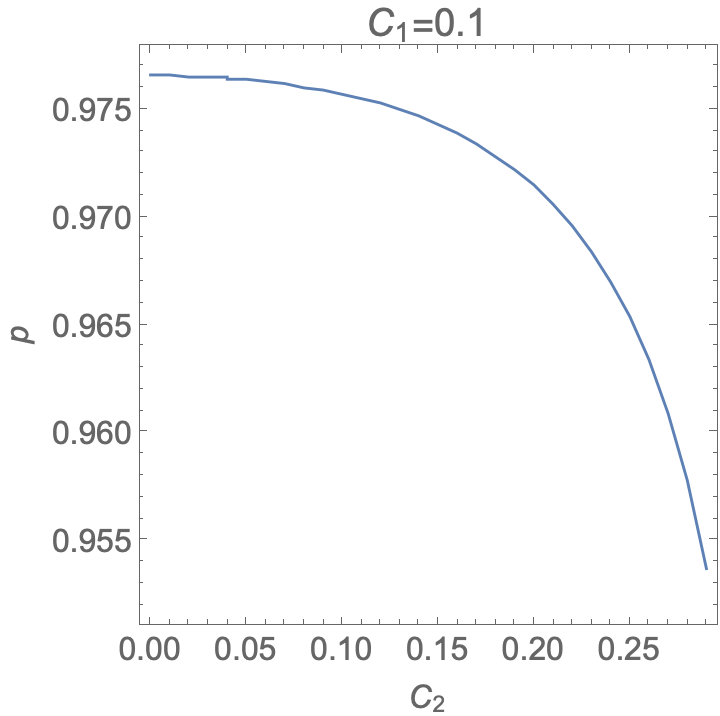}
\includegraphics[width=0.33\textwidth]{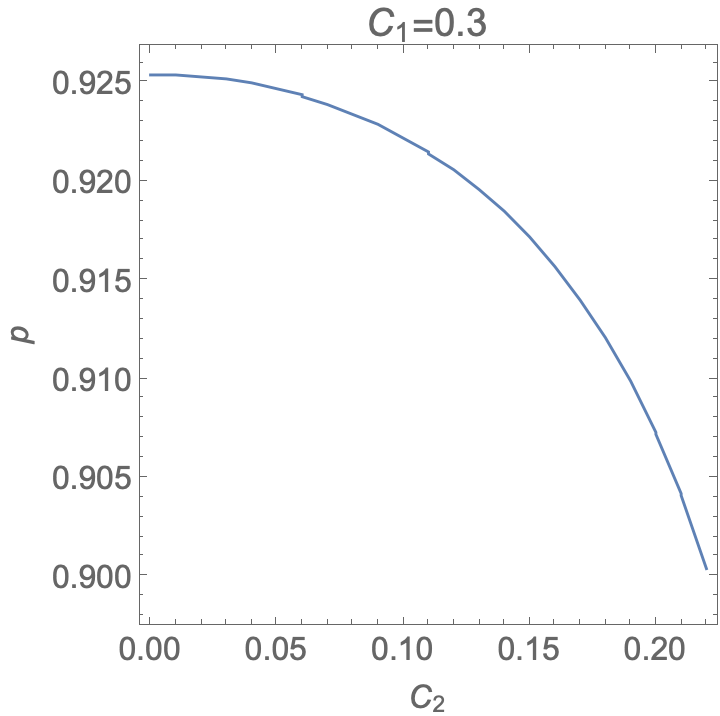}
\includegraphics[width=0.33\textwidth]{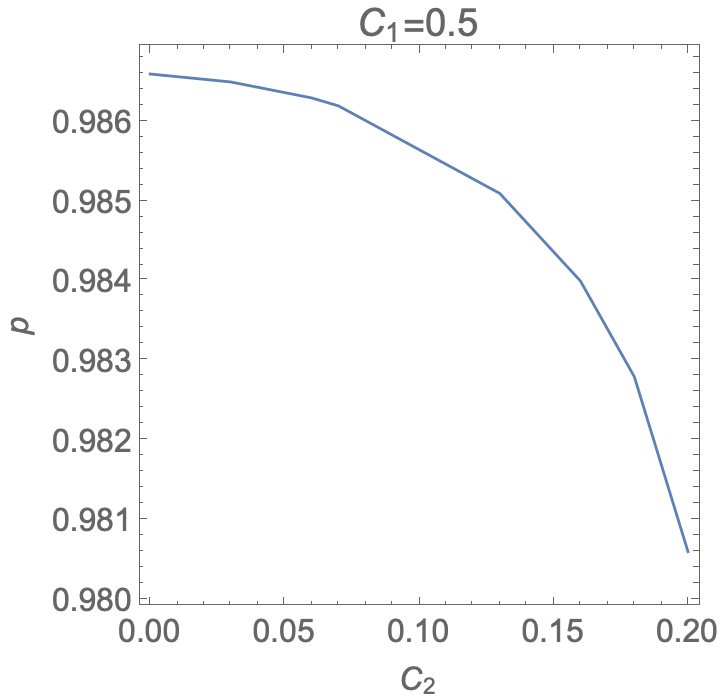}
\caption{ Radial power-law index $p$ as function of the spin parameter $C_2$ for different twist parameters.  Beyond the plotted values the index $p$ falls down precipitously.}
\label{C1} 
  \end {figure}

Corresponding solutions  for $p(C_1,C_2)$ are plotted in Figs. \ref{NoLC2}-\ref{C1}. Rotation first strongly modifies highly twisted solutions $p \rightarrow 0$,  Figs. \ref{NoLC2}. As the spin  parameter $C_2$ increase, the range of allowed values of $C_1-p$ decreases. Solutions become more dipolar-like (in terms of radial profile, $p\approx 1$). 
 
   %Rotating-twisted
   \begin{figure}
\includegraphics[width=0.99\linewidth]{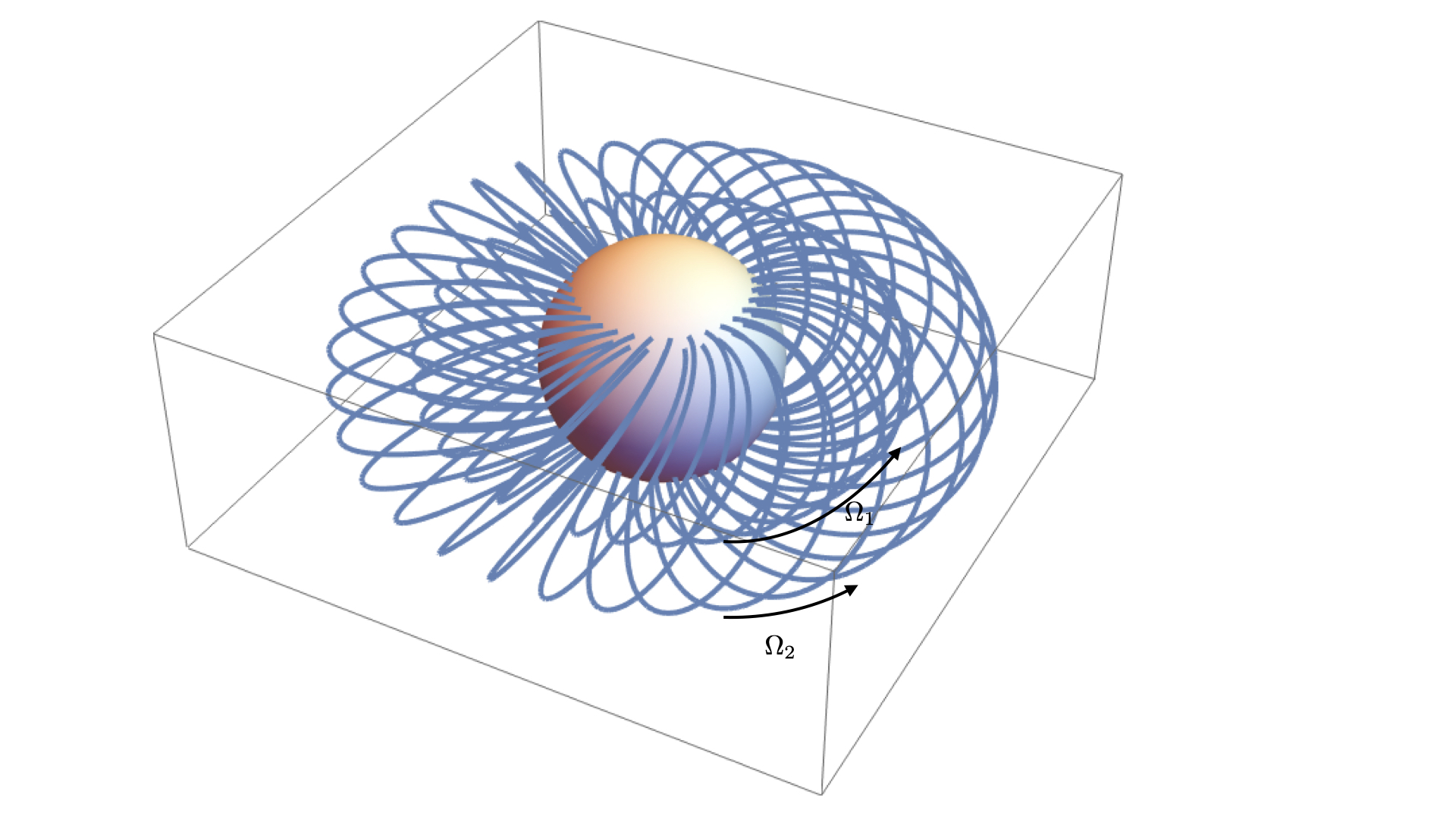}
\caption{Rotating and twisted \ms\ without \LC. Pictured are \Bf\ lines for two flux surfaces. On each flux  surface  the \Bf\ is twisted; in addition  flux surfaces are  rotating with spin frequency  $\Omega$ decreasing as a function of maximal radius (equivalently, becoming smaller for smaller   polar angles of the foot-points.}
\label{NoLC4} 
  \end {figure}

Note that in the cases when the  eigen-problem cannot be satisfied, there may   still  be solutions with no light cylinder, but those solutions will be non-self-similar. Eventually, for large enough spin the light cylinder will be formed.

 In the case of untwisted \mss, $C_1=0$, the maximal spin parameter is $C_2 = 0.34$, Fig. \ref{c1-00}. We note that finite twists $C_1 \neq 0$ allow for larger spins, Fig. \ref{C1}. 
 
 For small twists the \EM\ velocity is 
\be
\frac{\E \times \B}{B^2} \approx  \left\{-\frac{2 \sqrt{2} \sqrt{C_1} C_2 \sin ^6\theta  \cos \theta }{5+3 \cos (2 \theta
   )},-\frac{\sqrt{2} \sqrt{C_1} C_2 \sin ^7\theta }{5+ 3 \cos (2 \theta )},2 C_2 \sin
   ^3\theta \right\}
   \label{EMdrift}
   \ee
   It is independent of the radius and changes sign at the equator. The Pointing  flux   $\propto r^{-2 (2+p)}$ is zero at infinity: the pulsar does not spin down.

 For each solution the angular velocity decreases with radius and increases with polar angle
  \be
  \Omega= C_2 \left(1+\frac{1}{p} \right) \frac{F(\mu)^{1/p}}{r} = \frac{R_{NS}}{r}  {F(\mu)^{1/p}}  \Omega_0
  \label{Omega}
  \ee
where  typical value at the surface is 
  \be
  \Omega_0 = {C_2 \left(1+\frac{1}{p} \right) } \frac{c} {R_{NS}}
  \ee
  (each flux surface rotates as   a solid body.)
  
  Azimuthal velocity is 
  \be
  v_\phi =  \Omega r \sin \theta  =  (  R_{NS} \Omega_0)\sin \theta F^{1/p}
  \ee
  
  The locations of the \LC\ is determined by $v_\phi =c$.
  For the exemplary case of  $C_2=0.1, C_1 = 0.66, p = 0.79$ we have $\Omega_0 =0.227$; the maximal value of $\Omega r \sin \theta $  equals  $0.241 \leq 1$. There is no \LC.
  
\begin{figure}
\includegraphics[width=0.99\textwidth]{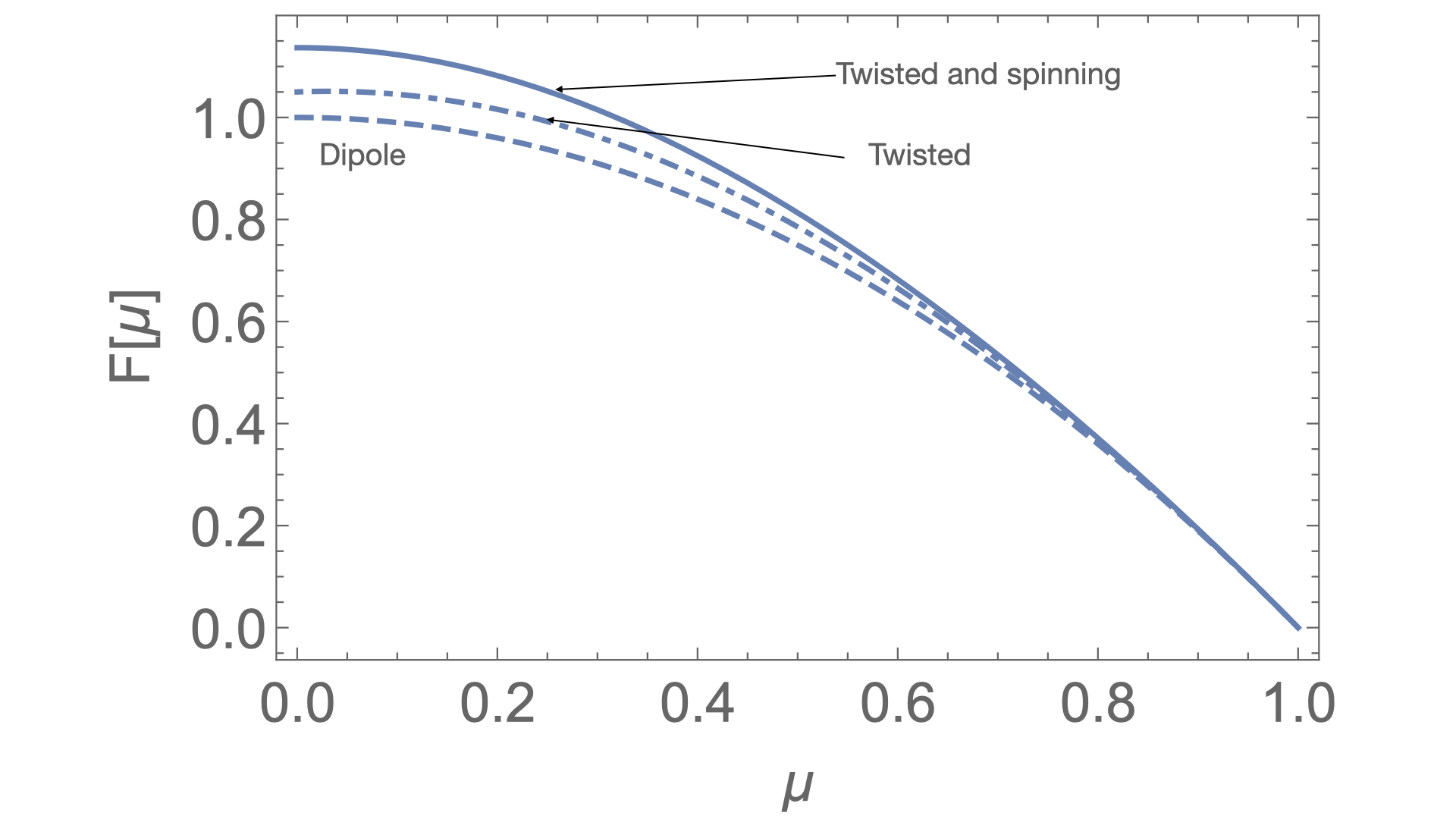}
\caption{  Example that both twist and rotation rotation leads to  ''inflation'' of field lines. Dashed line is for dipole $F=1-\mu^2$; dot-dashed is for twisted non-rotating case $C_2=0, C_1 = 0.66, p = 0.867$, solid line is rotating case  $C_2=0.1, C_1 = 0.66, p = 0.79$ (same $C_1$ as non-rotating).}
\label{NoLC3} 
  \end {figure}

   Generally, rotation leads to further  (in addition to twisting) ''inflation'' of field lines, Fig. \ref{NoLC3}. (We stress that inflation here occurs due to the rotation of a flux surface as a whole, not due to the shearing of the foot-points of  a given field line). 

The electric potential
\be
\Phi = C_2 r^{-(1+p)} F^{1+1/p}
\ee 
leads to distributed charge
\be
\rho_e = \Delta \Phi  \propto r^{-(3+p)}\approx  \frac{ \Omega_0 B_{NS}} {4\pi  c } \left(  \frac{R_{NS} }{r} \right)^{-(3+p)}
\ee
  \Ef,
\be
\E = - \nabla \Phi  \propto r^{-(2+p)}
\ee
Near the \NS\ the radial component  of the \Ef\ integrated over the surface gives the central  charge $Q_c$
\be
Q_c \approx \frac{  \Omega_0 B_{NS} R_{NS}^3}{c}
\ee
 Since for $r \rightarrow \infty$ $E_r \to 0$ faster than $1/r^2$, the total distributed charge equals the central charge.

\section{Simulation}

\subsection{PHAEDRA code}

PHAEDRA code \citep{2012MNRAS.423.1416P} is a pseudo-spectral code developed specifically to study highly magnetized plasma regime, force-free electrodynamics, the vanishing-inertia limit of magnetohydrodynamics \citep{Gruzinov99,2006MNRAS.367...19K}

% Results from standard MHD codes can become inaccurate, for large plasmaâ magnetization is large, because of numerical error in the electromagnetic energy density.

The code solves  Maxwell's equations
\begin{equation}
\begin{array}{l}
\frac{1}{c} {\partial _t\mathbf{B}}=-\nabla \times \mathbf{E} \\
 \frac{1}{c} {\partial _t \mathbf{E}}=\nabla \times \mathbf{B}-\frac{4 \pi}{c} \mathbf{J}
\end{array}
\end{equation}
supplemented by the  force-free expression.
\begin{equation}
\rho \mathbf{E}+\mathbf{J} \times \mathbf{B}=0
\end{equation}
The force free Ohm's law is then 
\begin{equation}
\frac{4 \pi \mathbf{J_{FF}}}{c}=\frac{B \cdot \nabla \times \mathbf{B}-\mathbf{E} \cdot \nabla \times \mathbf{E}}{B^{2}} \mathbf{B}+\nabla \cdot \mathbf{E} \frac{\mathbf{E} \times \mathbf{B}}{\mathbf{B}^{2}}
\end{equation}
Condition $\mathbf{E}.\mathbf{B} =0$ is satisfied by construction, while $\mathbf{B}^2 - \mathbf{E}^2  \geq 0$ is assumed (and checked at each step).
The code uses a thin frictional absorbing layer next to the outer boundary. The code applies two spectral filters of  8th and 36th order, to maintain stability \citep{2012MNRAS.423.1416P}.

Our results further indicate that the code is very stable and efficient,  and has low numerical dissipation. The full numerical grid is defined in axisymmetric spherical coordinates and composed of  $N_{r} \times N_{\theta} $ cells along the radial and  $\theta$ directions, respectively. The simulation zone in our work extends from the stellar surface $r_{min} =R_{NS}$ up to $r_{max} =100R_{NS}$.

The inner boundary is set as the radius of the star ($r=R_{NS}$) and the following conditions are strongly enforced at every Runge-Kutta substep.
\begin{equation}
\begin{array}{l}
B_{r}=B_{r}(\theta)\\
 E_{\theta}=-\Omega B_{r} \sin \theta\\
 E_{\phi}= 0
 \end{array}
\end{equation}

\subsection{Initialization}
 
 Analytical self-similar solutions, \S \ref{Self-similar}, are approximations: the system is generally non-self-similar (there is a special scale - the \LC). Thus, we do not expect the numerics to match exactly with analytics, just to follow the general trend. With this in mind, 
we initialize the simulation setup with analytical approximation for the structure of non-rotating twisted magnetospheres for small/mild twist parameter $C_{1}$, {\it and} the values of $C_2$  from the self-similar solution, Fig.  \ref{NoLC2}.   For small twists we can find 
analytical relation for the structure of non-rotating \ms\  by expanding near $p = 1,\, C_{1}  = 0 $ and the dipolar flux function \citep{2013arXiv1306.2264L}:
\ba &&
p = 1 - 8  C_{1}/35
\nn &&
{ F} =
(1- \mu^2) \left( 1 + {1 \over 140} (1 -  \mu^2)( 17 - 5  \mu^2) C_{1} \right)
\nn &&
\mu = \cos \theta
\label{F}
\ea
In this approximation the  flux 
magnetic field components and twist angle are given by
\ba &&
\frac{B_r }{B_{NS}} =  { 2  \mu \over  \tilde{r}^{2+p}} \,
\left(1+ {3 (13 - 18 \mu^2 +5 \mu^4)  \over 140} C_{1} \right)
\nn &&
\frac{B_\theta }{B_{NS}}  =  p { \sqrt{ 1 -  \mu^2}  \over  \tilde{r}^{2+p}}
\, \left(1 + {(17 -  22 \mu^2 + 5 \mu^4) C_{1} \over 140}\right)
\nn &&
\frac{B_\phi }{B_{NS}}  = 
{ \sqrt{ \frac{ C_{1} p}{1+p} } \frac{ (1 -  \mu^2)^{3/2} } { \tilde{r}^{2+p }}}
\nn &&
\frac{j_r}{B_{NS}/{4\pi} }  = 2 \sqrt{2} \mu (1-\mu^2)  \tilde{r}^{-3-p } \sqrt{C_{1}}  
\nn &&
\frac{j_\theta}{B_{NS}/{4\pi} } = \sqrt{2}  (1-\mu^2)^{3/2}  \tilde{r}^{-3-p } \sqrt{C_{1}} 
\nn && 
\frac{j_\phi}{B_{NS}/{4\pi} }  =  (1-\mu^2)^{5/2}  \tilde{r}^{-3-p } C_{1} 
\nn &&
\Delta \phi =  \sqrt{2 C_{1}}  \mu_{fp}
%\nn &&
%L_{sd} \sim B^2_{NS} R_{NS}^2 c \left(\frac { \Omega  R_{NS} }{ c}\right)^{4-16 C/35}
\nn &&
 \tilde{r} = \frac{r}{R_{NS}}
\label{small}
\ea
where $C_{1}$ is  a twist  parameter,  $ \mu_{fp}$ is the cosine of the polar angle of the northern foot point and $B_{NS}$ is the equatorial \Bf.
 
  \subsection{Results}

%Results of numerical simulations are  in agreement with the analytics, Figs. \ref{00} and \ref{3}. 

\begin{figure}
\includegraphics[width=0.99\textwidth]{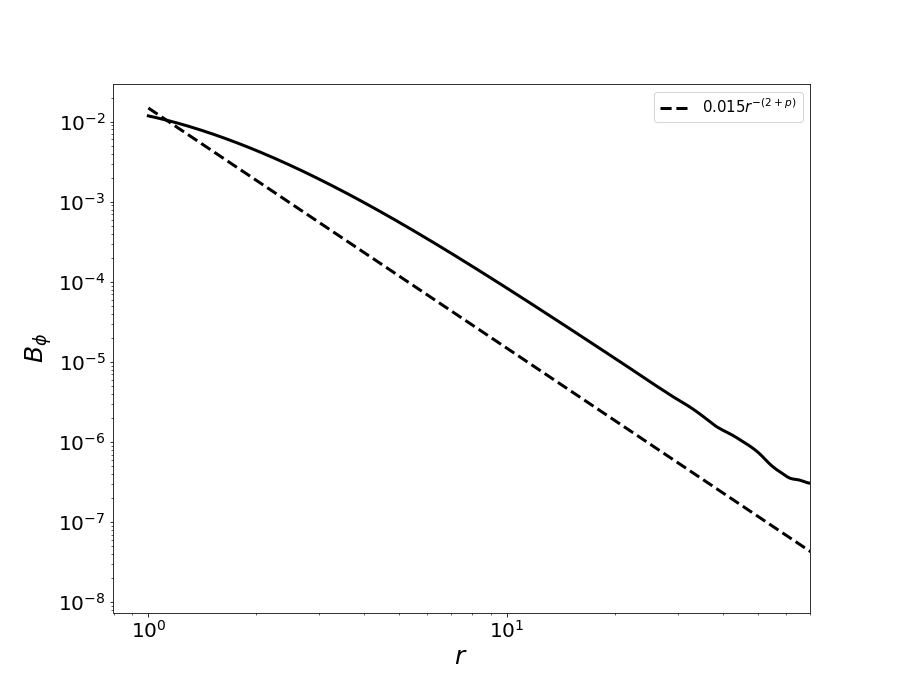}
\caption{Time slice of toroidal \Bf\ $B_{\phi}$ near the equator for  $C_{2}=0.10, C_{1} = 0.005, p = 0.9989$.  Solid line is our simulation results, dashed line is the theoretical expectation  $r^{-(2+p)}$.}
\label{0} 
  \end {figure}

In order to demonstrate that the simulated magnetic field follows the radial scaling predicted by analytics, we plot a time slice of toroidal magnetic field $B_{\phi}$ at $\theta=\frac{\pi}{2}$ in Fig. \ref{0} and try to fit it to $r^{-(2+p)}$. We see that the our radial scaling is in agreement with the analytics. 
  
\begin{figure}[htb!]
 \centering
 \rotatebox{90}{$  \hspace{3em}  C_{1}=0.005,C_2=0.10$}
\includegraphics[width=0.25\textwidth,height=0.22\textheight]{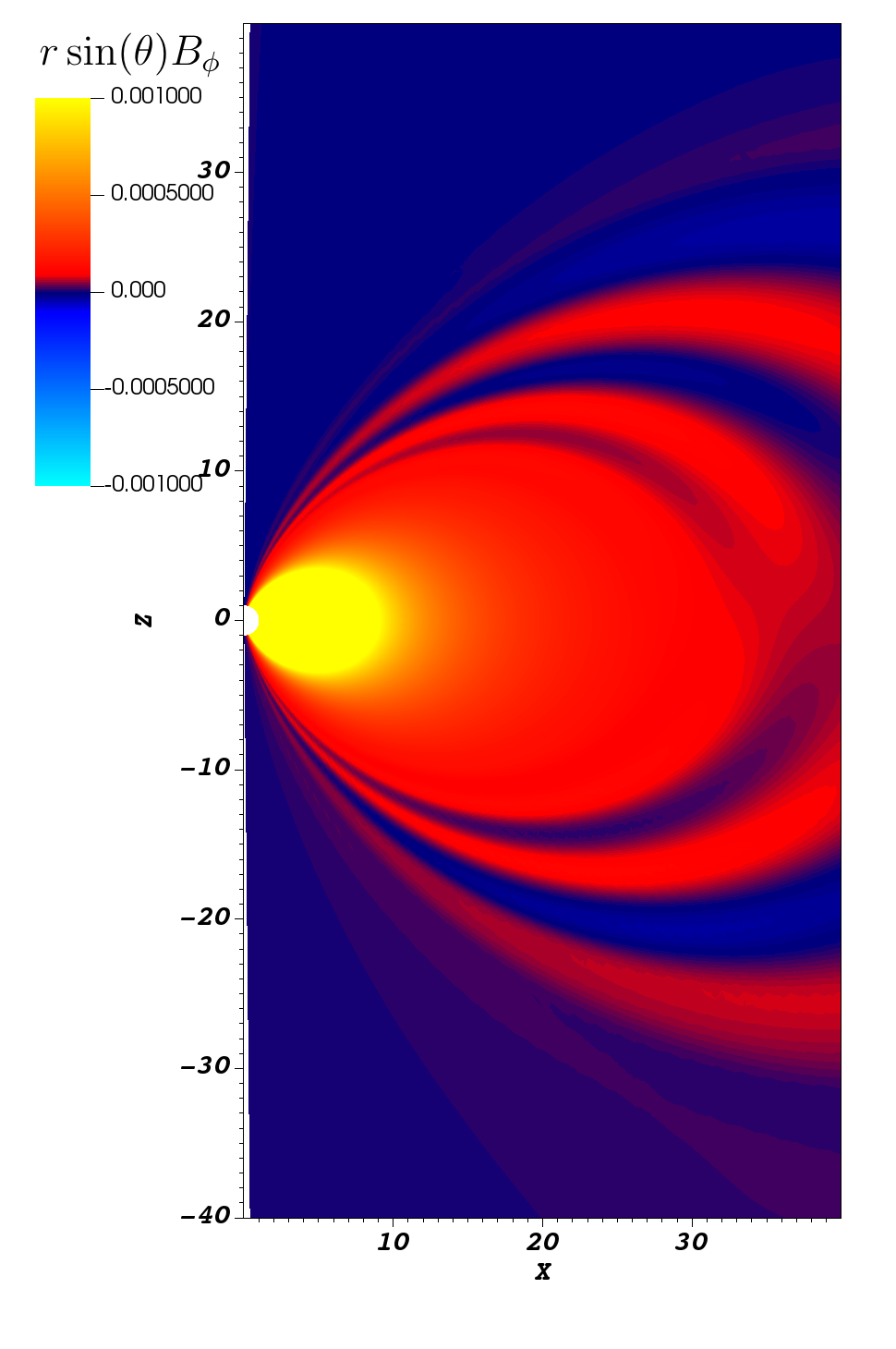}
\includegraphics[width=0.25\textwidth,height=0.22\textheight]{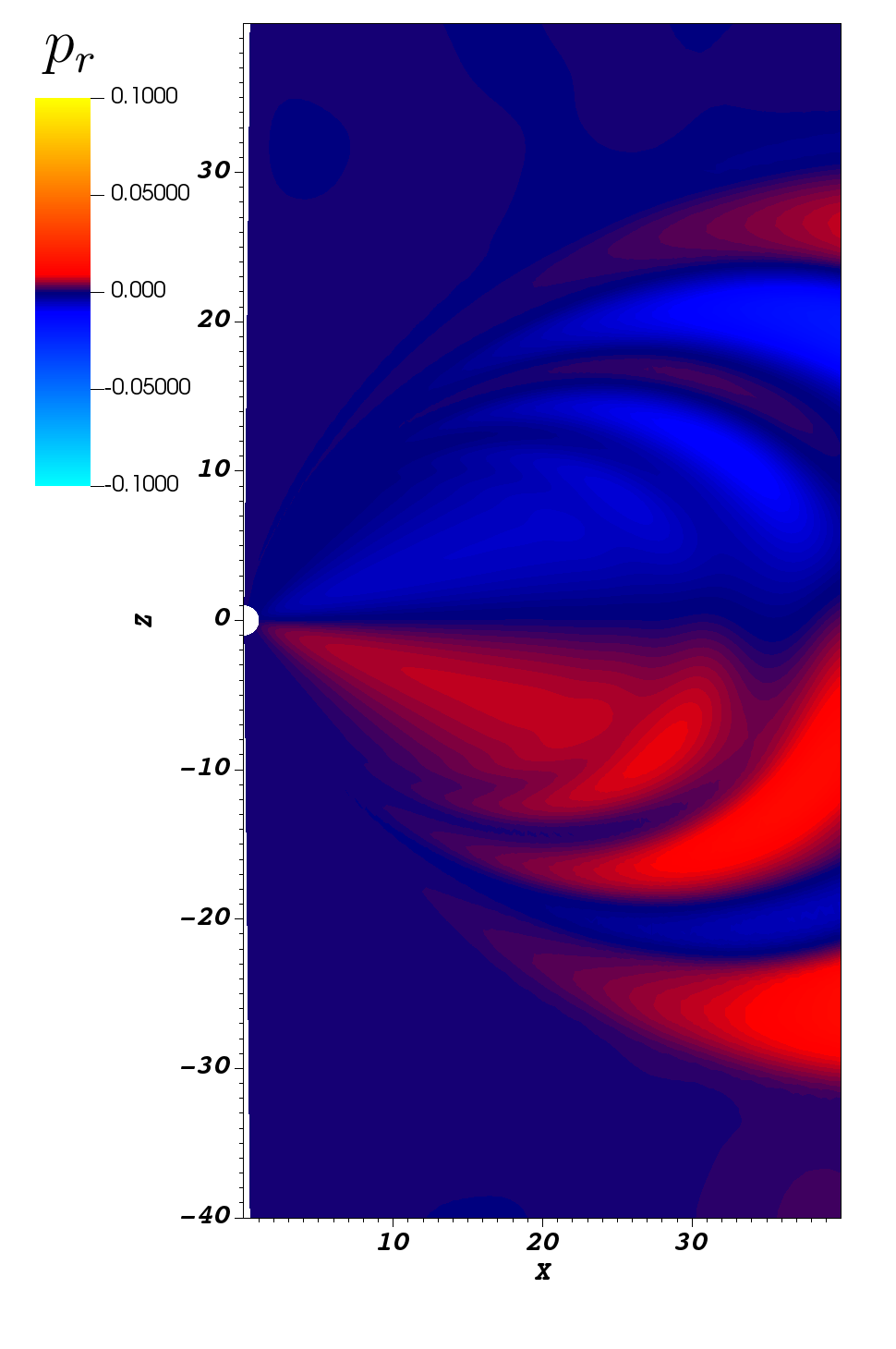}
\includegraphics[width=0.25\textwidth,height=0.22\textheight]{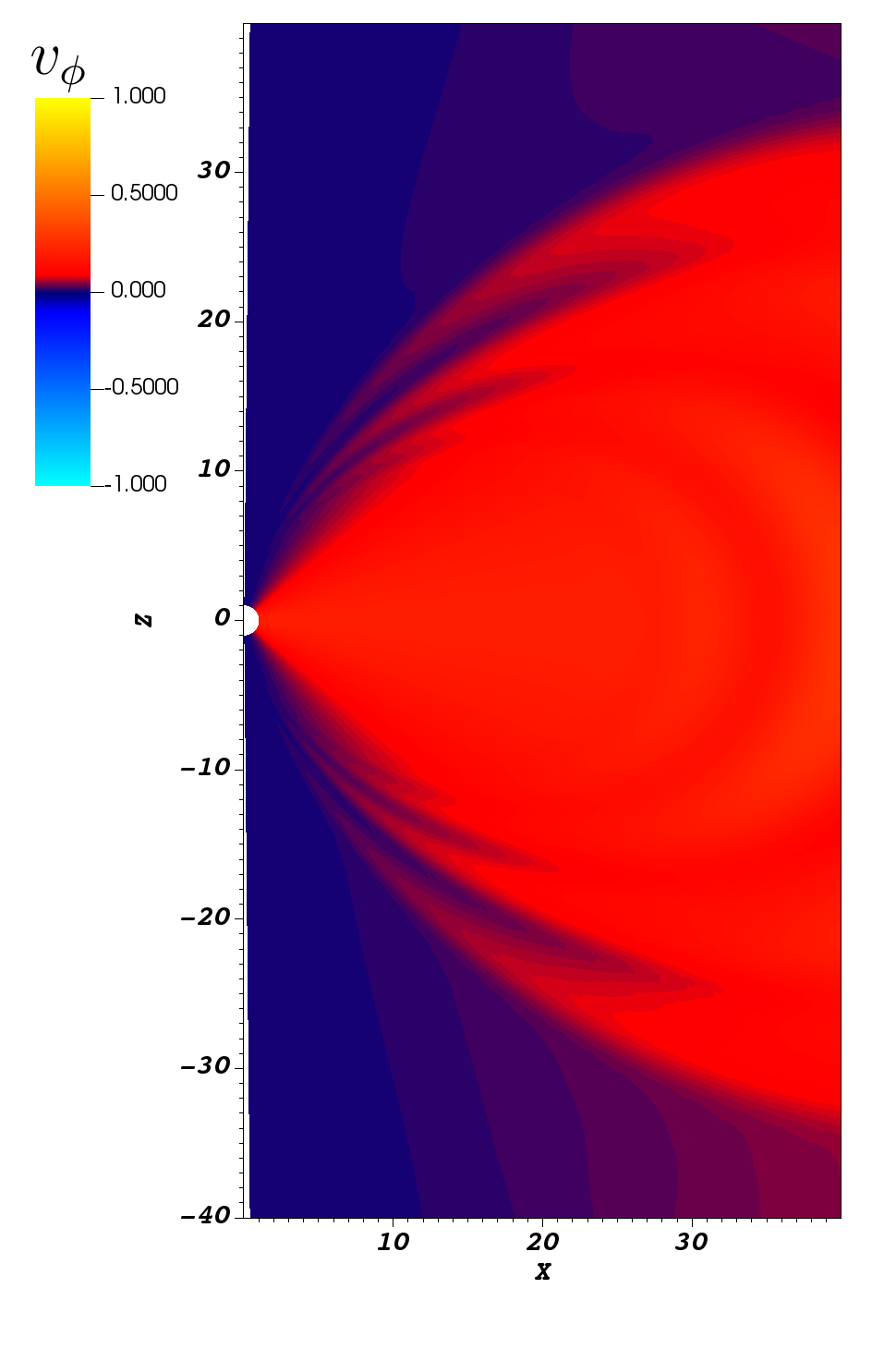}\\
\rotatebox{90}{$  \hspace{3em}  C_{1}=0.66,C_{2}=0.10$}
\includegraphics[width=0.25\textwidth,height=0.22\textheight]{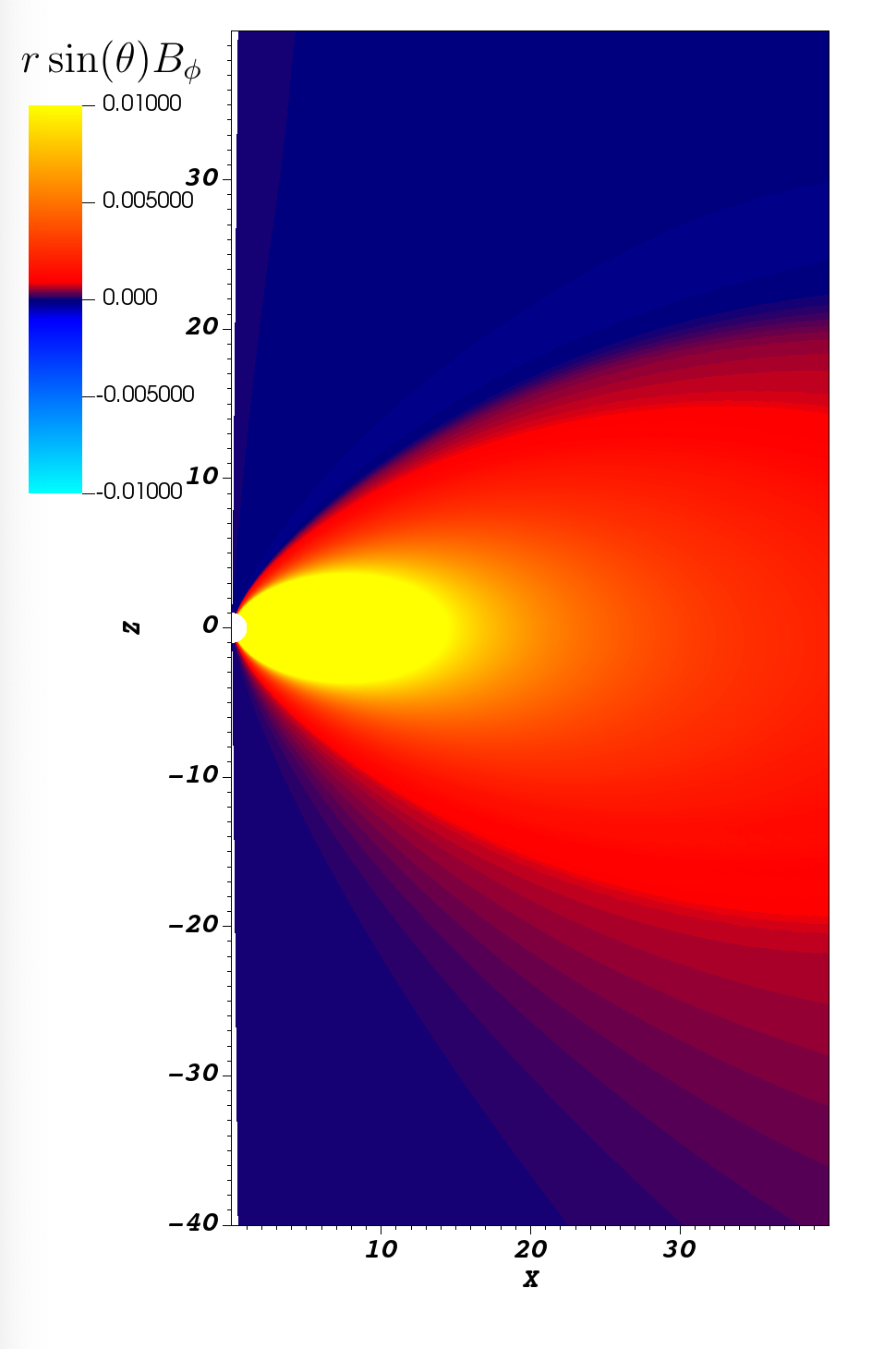}
\includegraphics[width=0.25\textwidth,height=0.22\textheight]{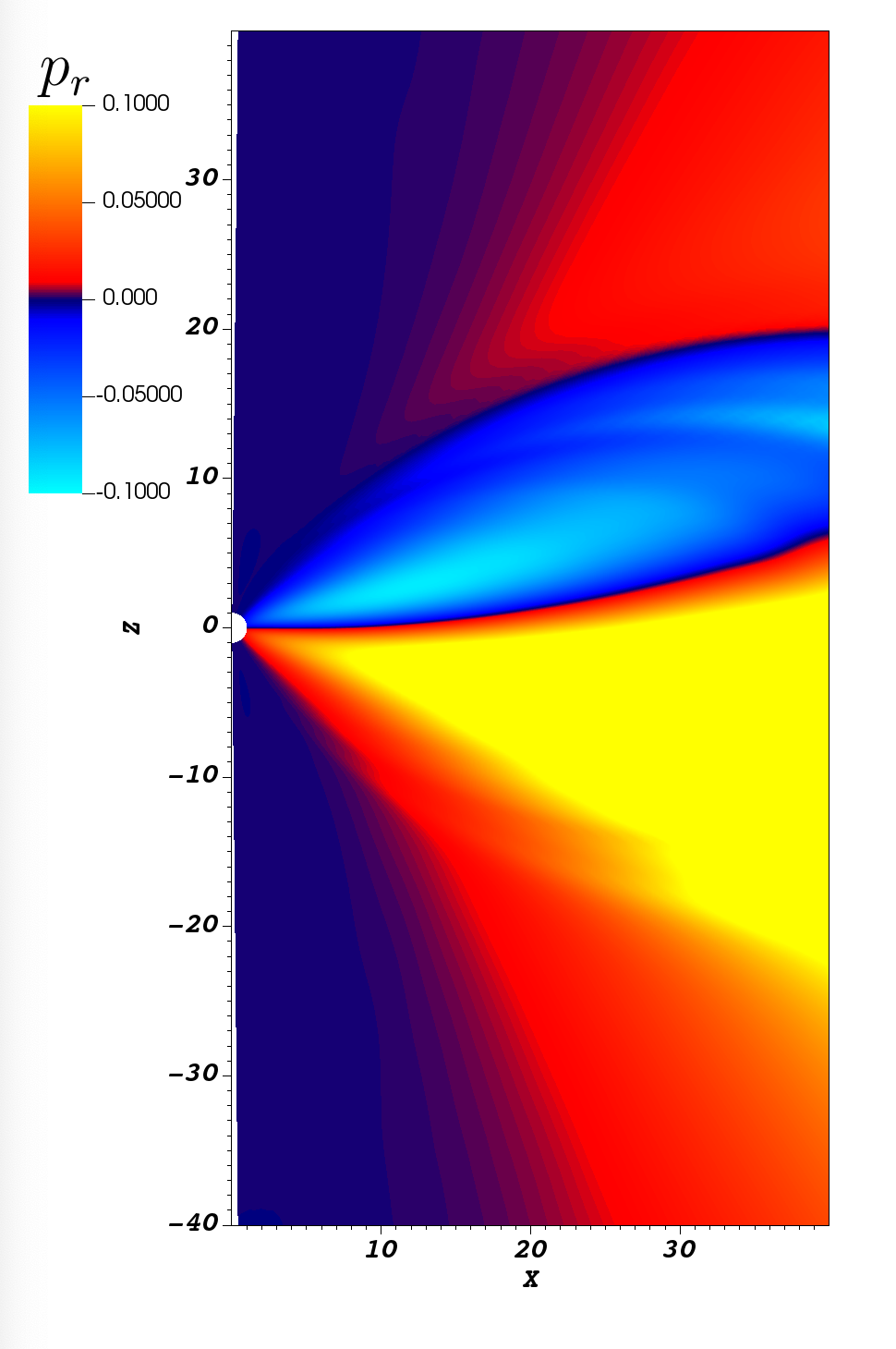}
\includegraphics[width=0.25\textwidth,height=0.22\textheight]{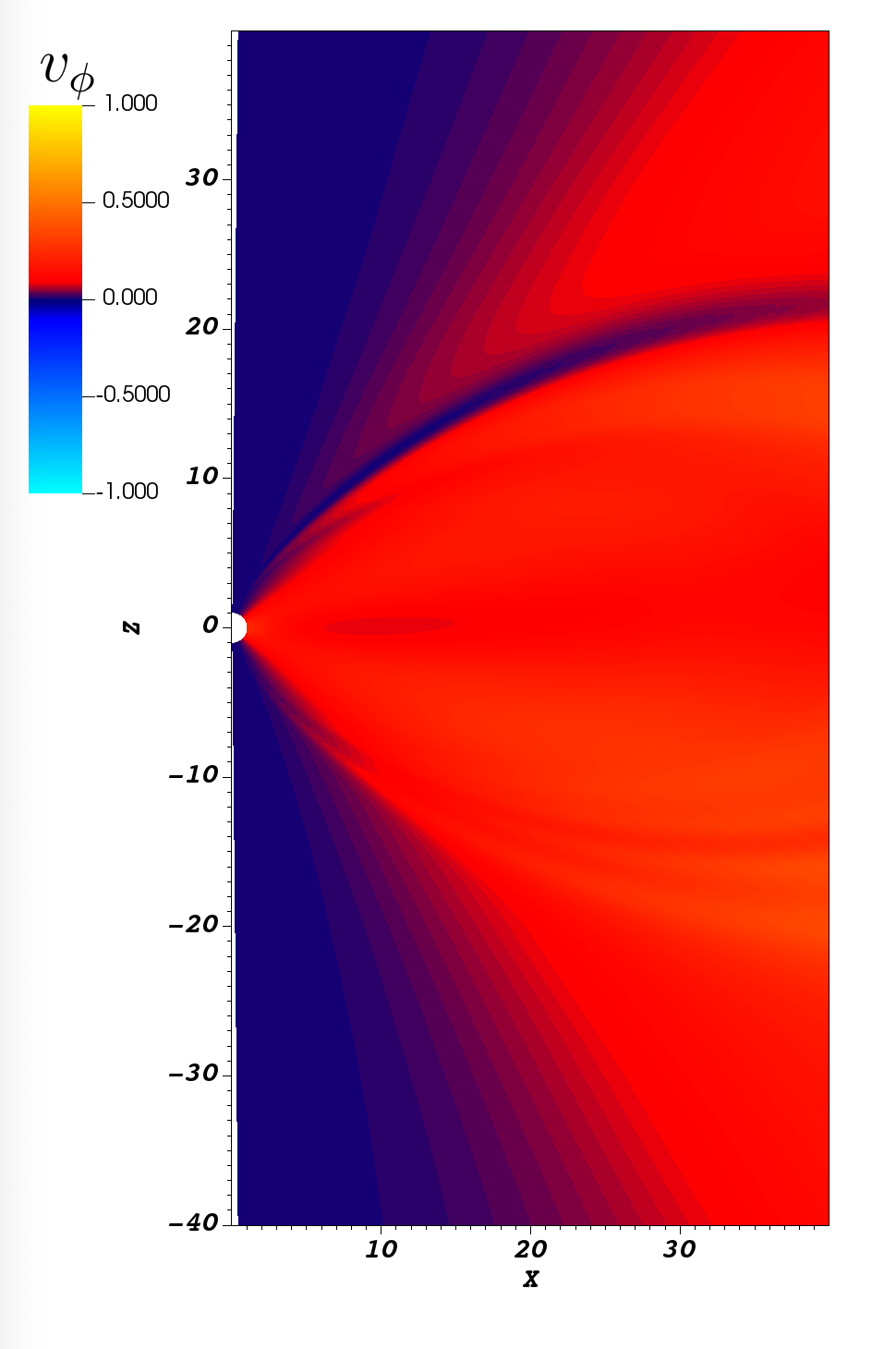}\\
\rotatebox{90}{$  \hspace{3em}  C_{1}=0.005,C{2}=0.20$}
\includegraphics[width=0.25\textwidth,height=0.22\textheight]{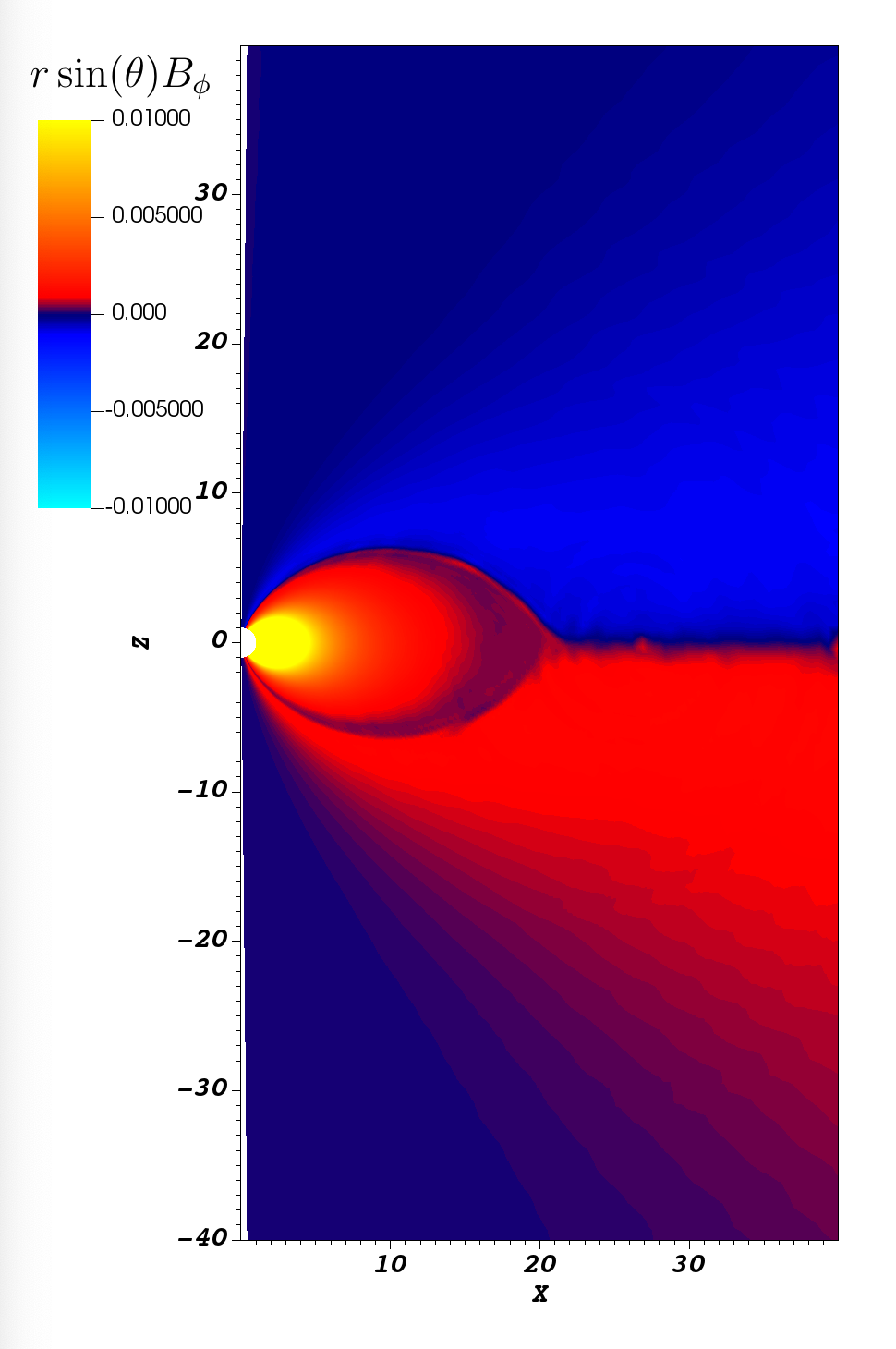}
\includegraphics[width=0.25\textwidth,height=0.22\textheight]{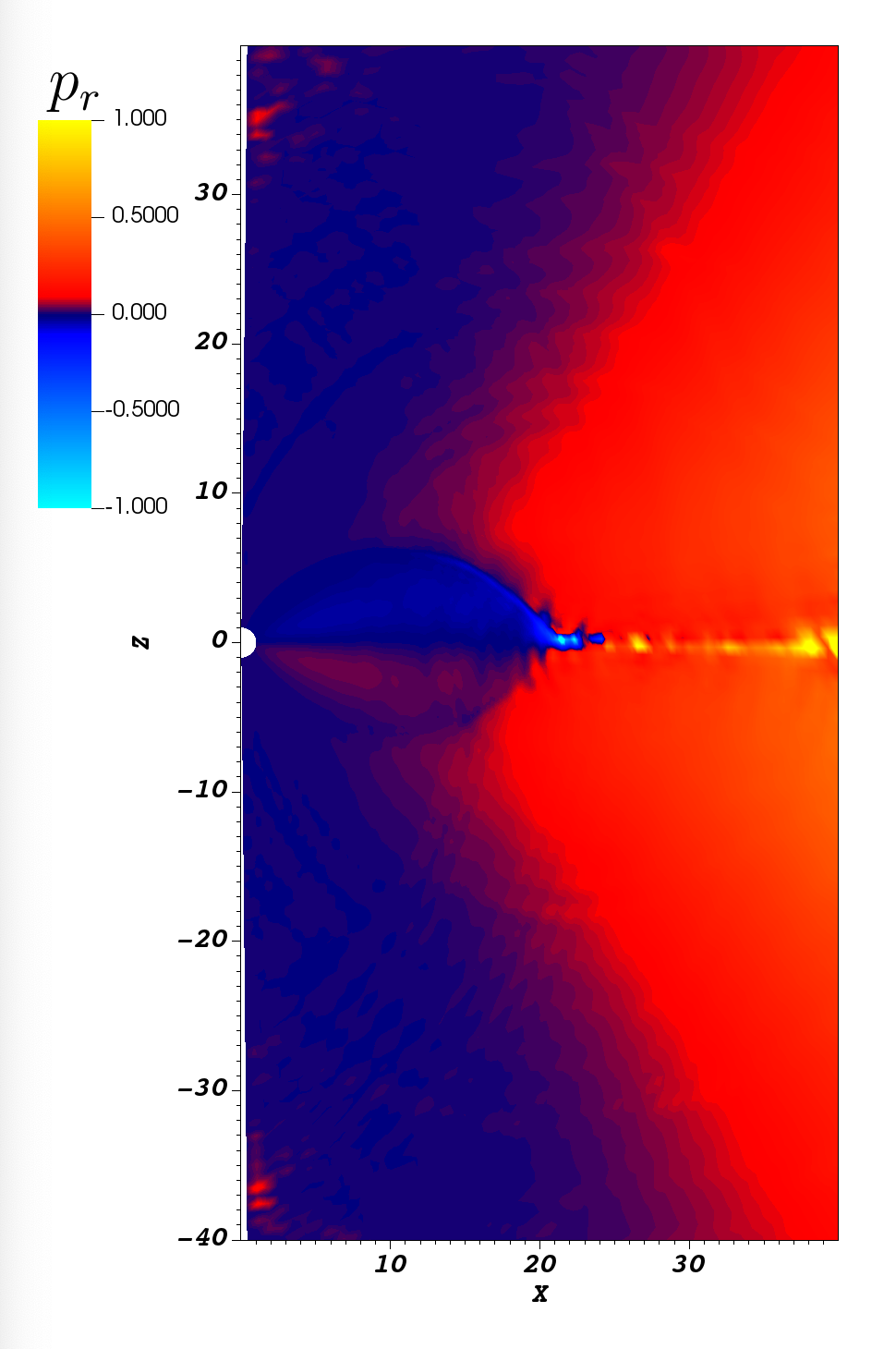}
\includegraphics[width=0.25\textwidth,height=0.22\textheight]{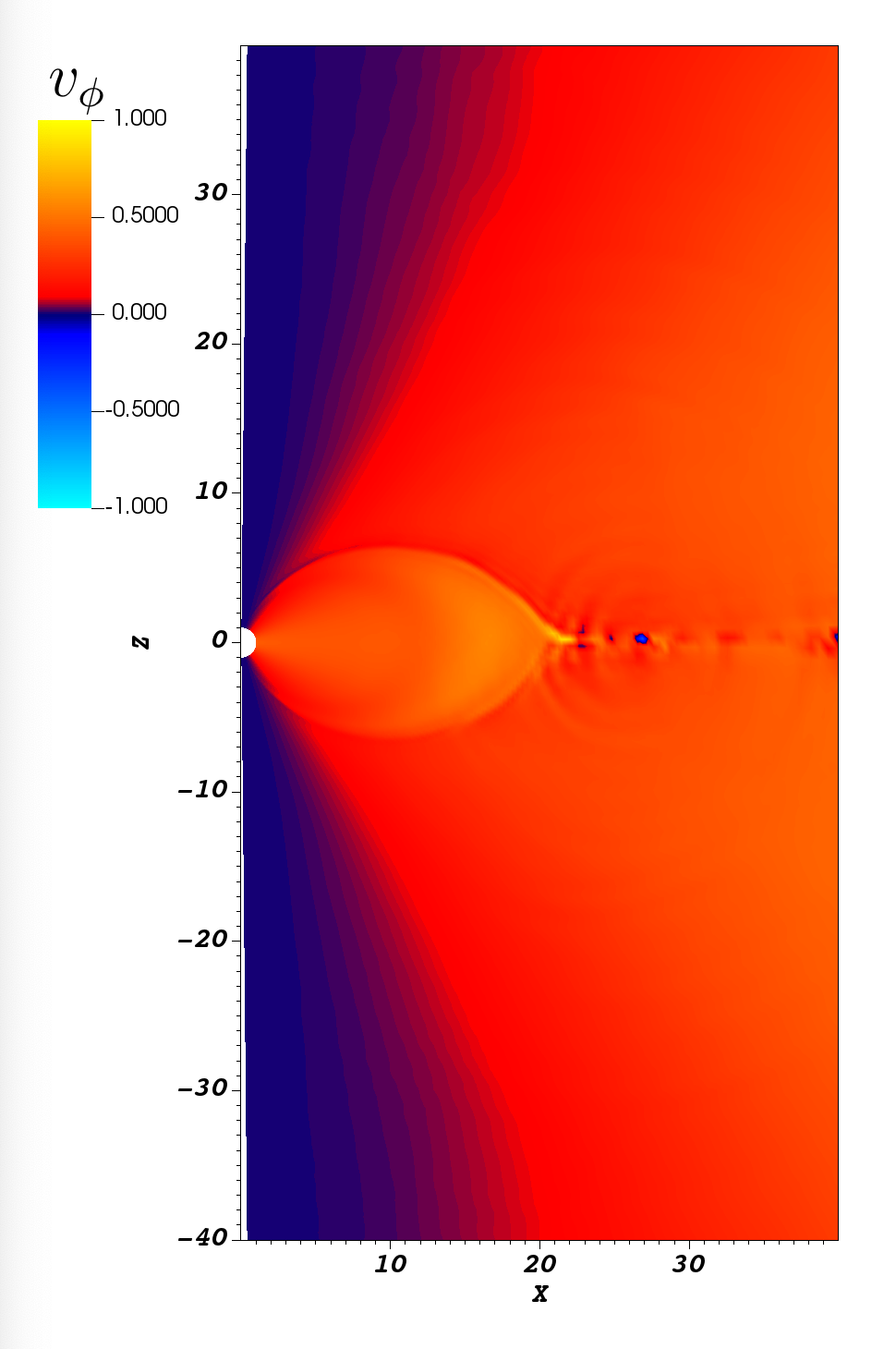}
\caption{ Zoomed-in time slice of our force-free simulation depicting toroidal \Bf\ $r \sin(\theta)B_{\phi}$, radial momentum $p_r$ and toroidal velocity $v_\phi$ for the fiducial \ms\ without \LC. In top panel we plot $C_1 = 0.005,C_2=0.10, p = 0.9989$ case, middle panel is for $C_1 = 0.66,C_2=0.10, p = 0.79$, bottom panel is for $C_1 = 0.05,C_2=0.20, p = 0.9854$. The radial momentum is the $\E\times \B \propto C_2 \sqrt{C_1}$ drift, Eq. (\protect\ref{EB}).  Since $\Omega \propto 1/r$, Eq.  (\protect\ref{Omega}), azimuthal velocity $v_\phi$  remains approximately constant with radius, Eq. (\ref{EMdrift}), top right panels. 
We don't observe  \LC\ in top two rows. At higher spin  parameter $C_2=0.2,C_{1}=0.05,p=0.9854$ (lower row) the light cylinder can be clearly identified by the location of the Y-point. The wind is generated (this is especially clearly seen in radial momentum $p_r$, central panel. }
\label{00} 
  \end {figure}

In Fig. \ref{00} we plot the scaled azimuthal magnetic field, radial momentum, and the azimuthal velocity for three sets of $C_{1},C_{2}$. For $C_1 =0.005 \sim0 $ as evident in the top panel, the structure of the rotating \ms\ clearly matches the analytical result: there are no open field lines, not radial Poynting flux, no energy losses. We see  similar structure for the case of $C_2=0.1, C_1 = 0.66, p = 0.79$  (twisted sheared configuration) in bottom panel. 

\begin{figure}[!h]
\centering
 \rotatebox{90}{$  \hspace{3em}  C_{2}=0.2$}
\includegraphics[width=0.22\textwidth,height=0.22\textheight]{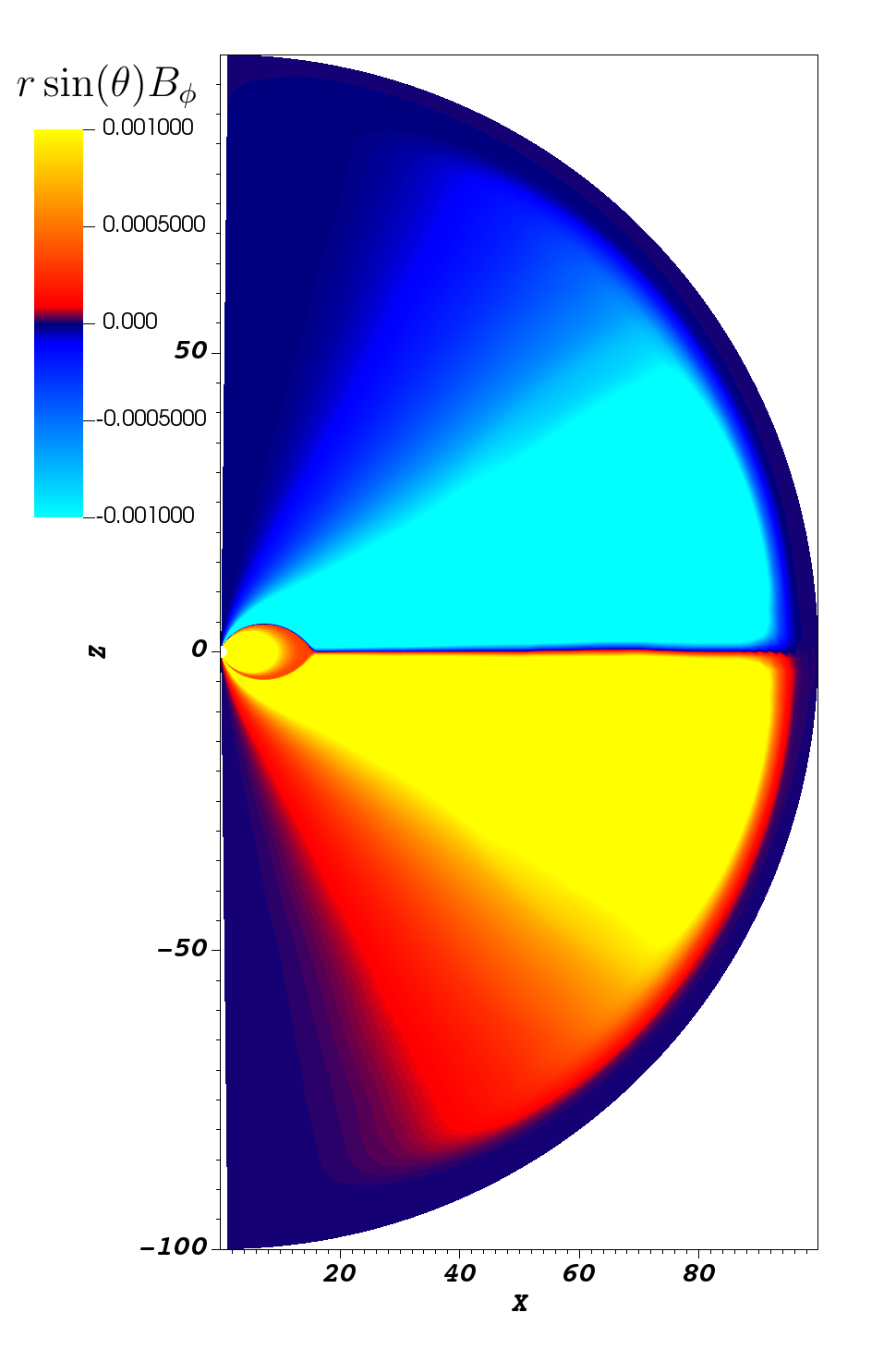}
\includegraphics[width=0.22\textwidth,height=0.22\textheight]{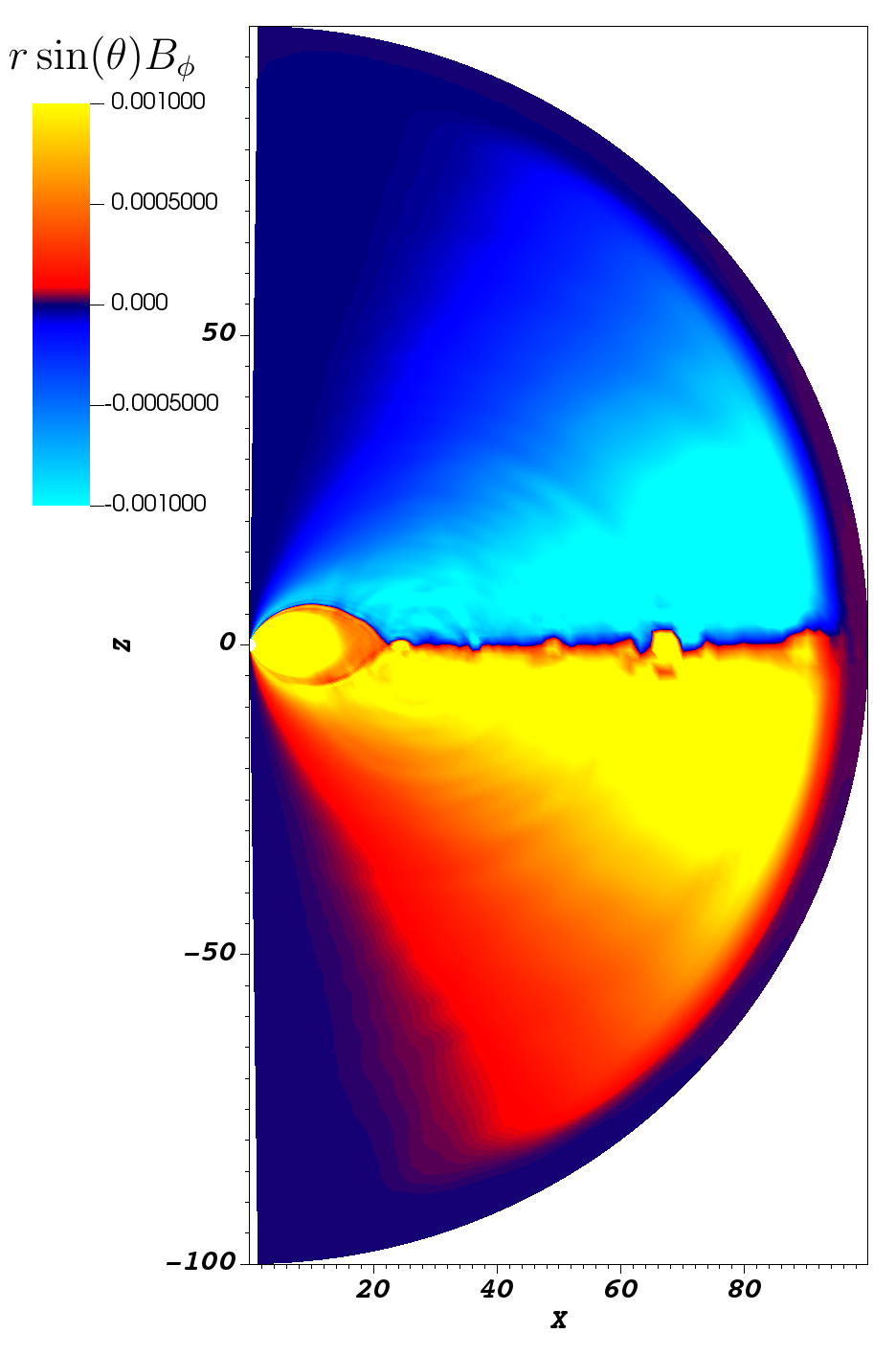}
\includegraphics[width=0.22\textwidth,height=0.22\textheight]{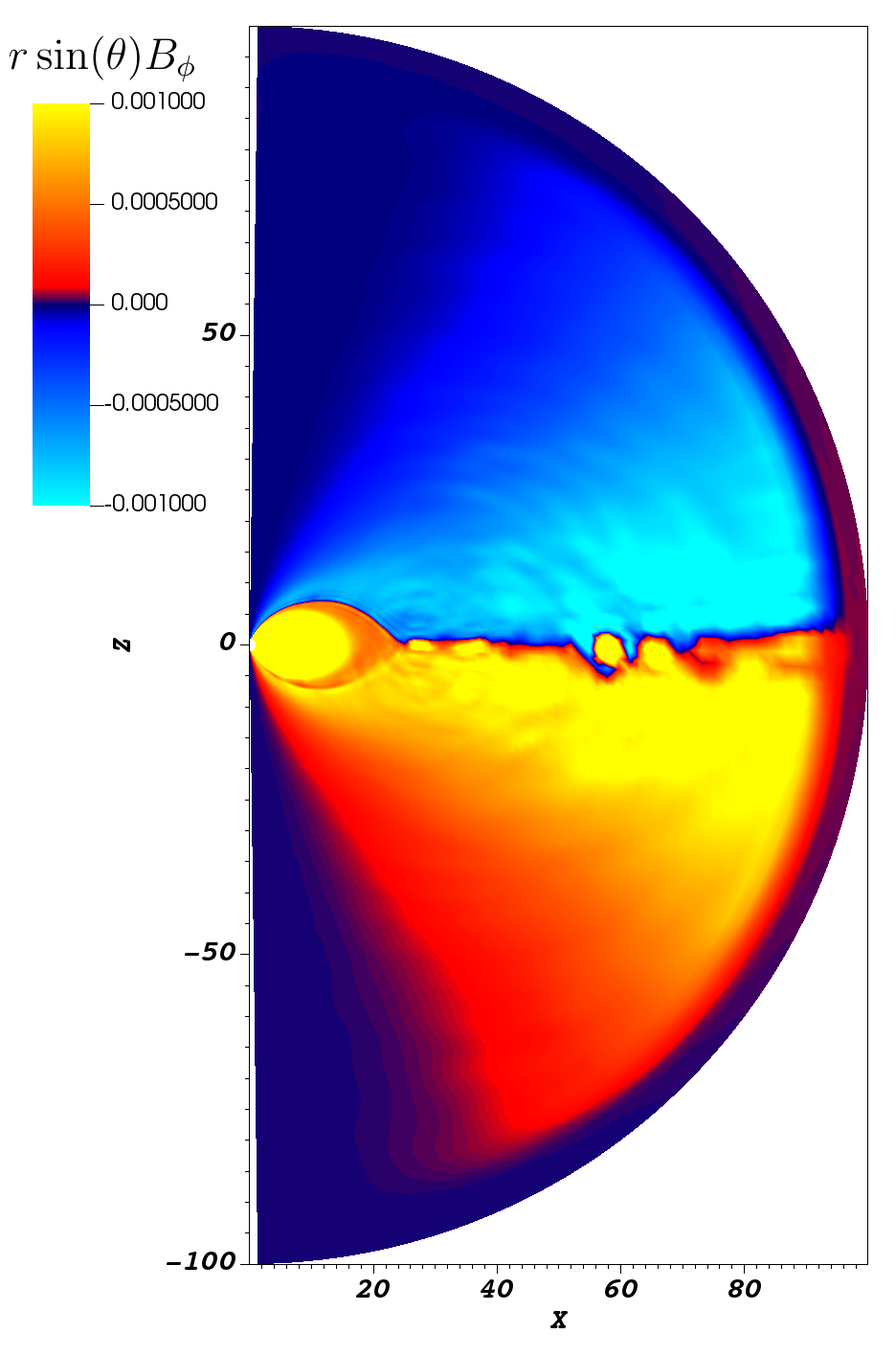}
\includegraphics[width=0.22\textwidth,height=0.22\textheight]{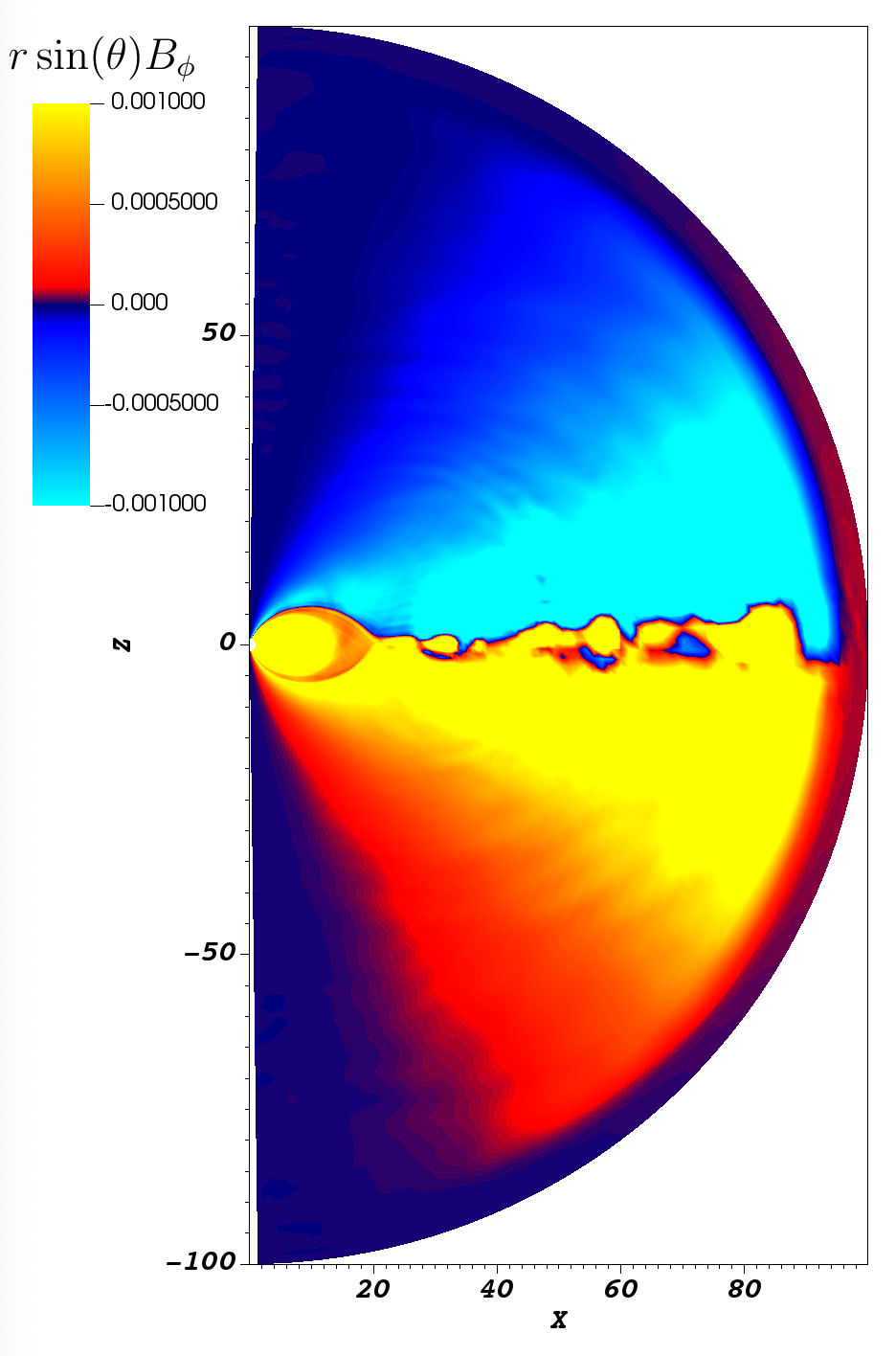}\\
\rotatebox{90}{$ \hspace{3em} C_{2}=0.15$}
\includegraphics[width=0.22\textwidth,height=0.22\textheight]{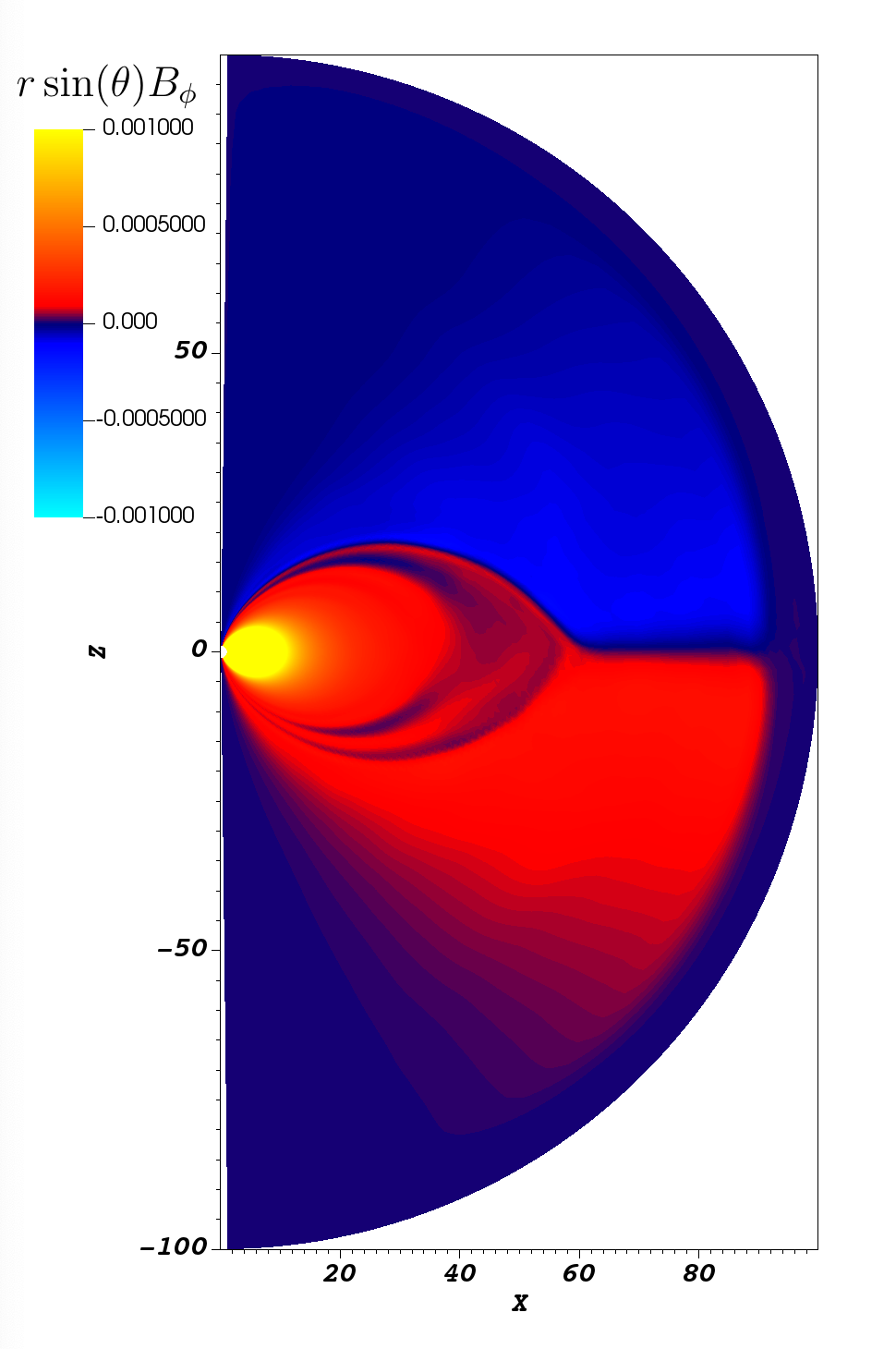}
\includegraphics[width=0.22\textwidth,height=0.22\textheight]{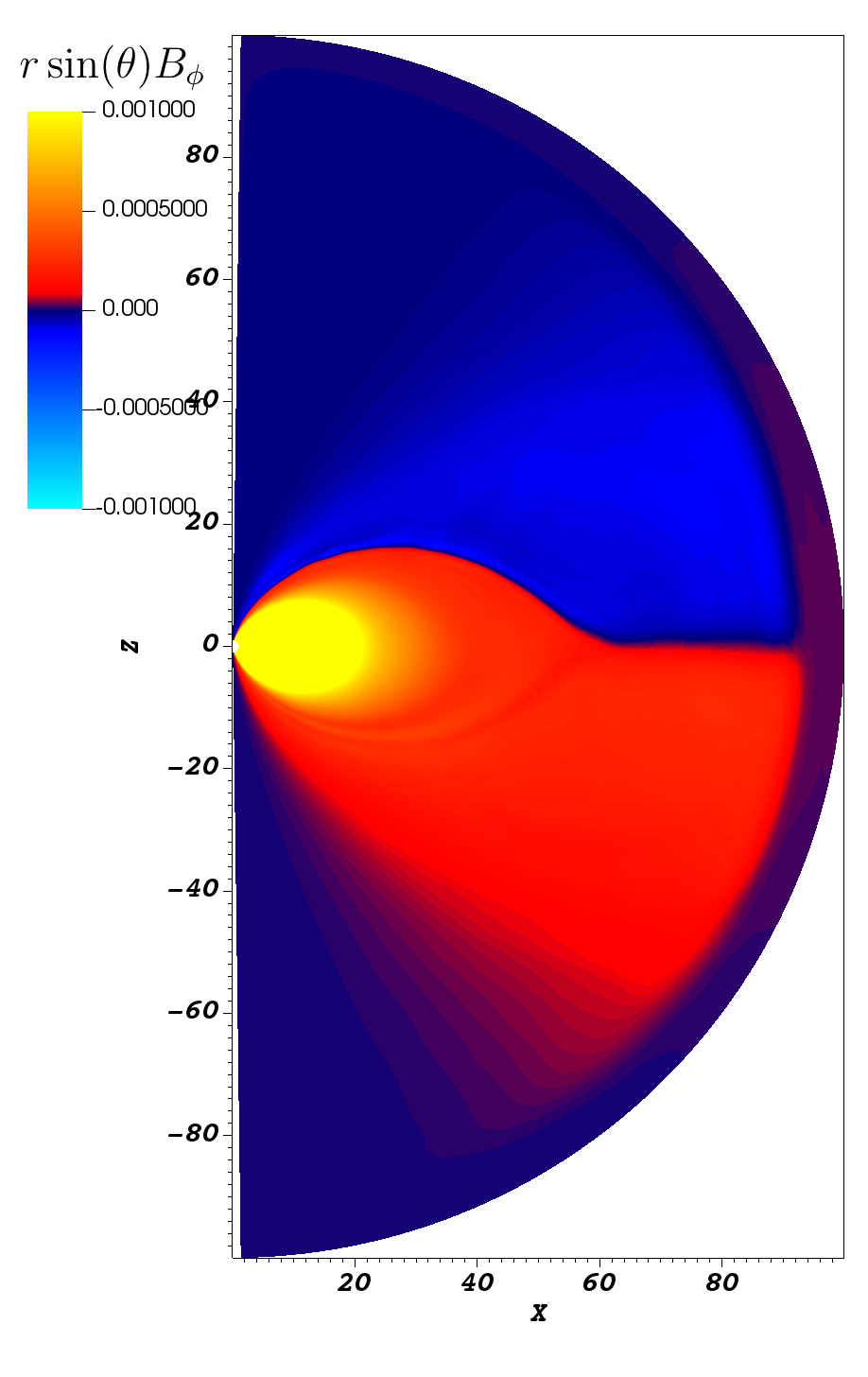}
\includegraphics[width=0.22\textwidth,height=0.22\textheight]{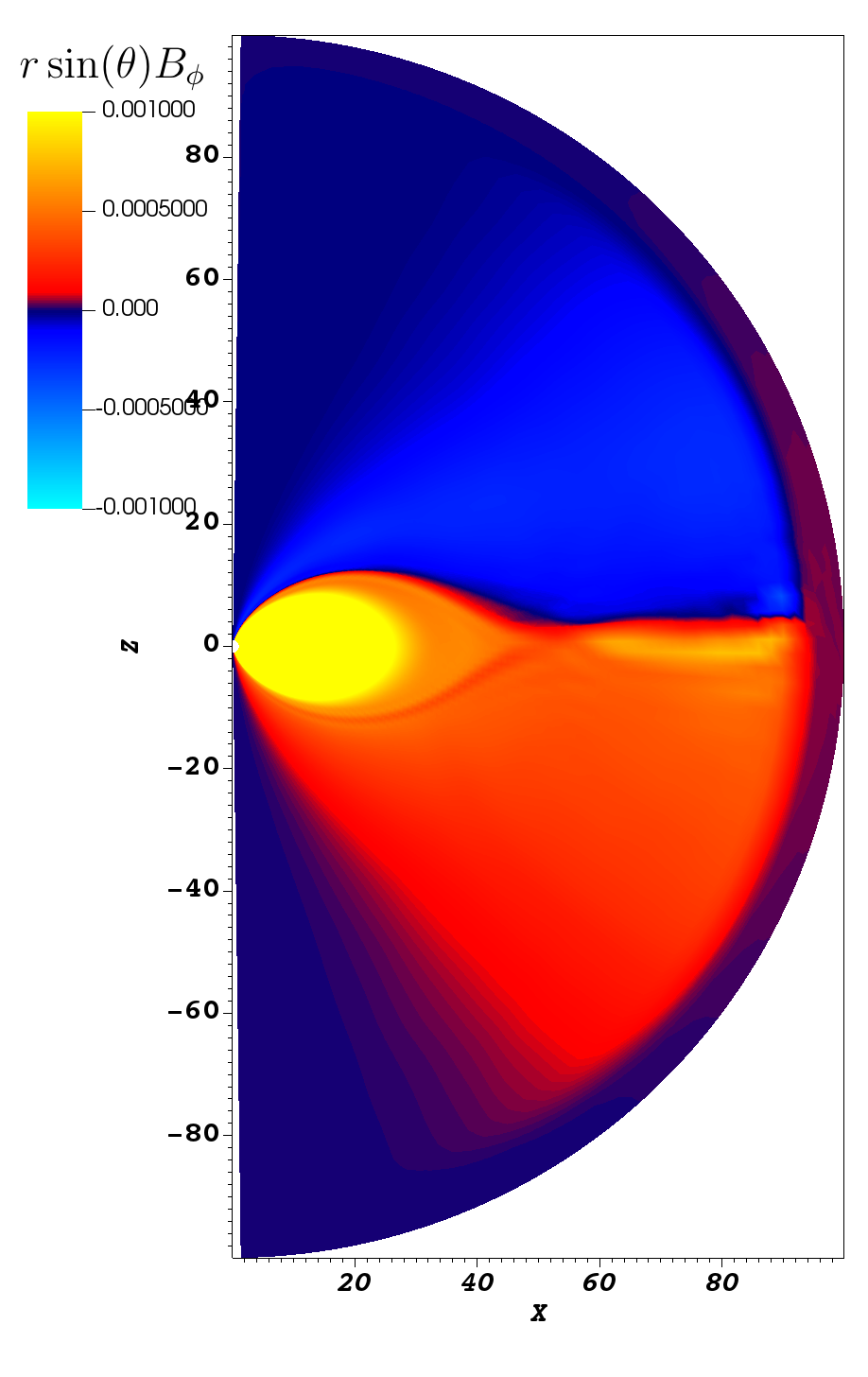}
\includegraphics[width=0.22\textwidth,height=0.22\textheight]{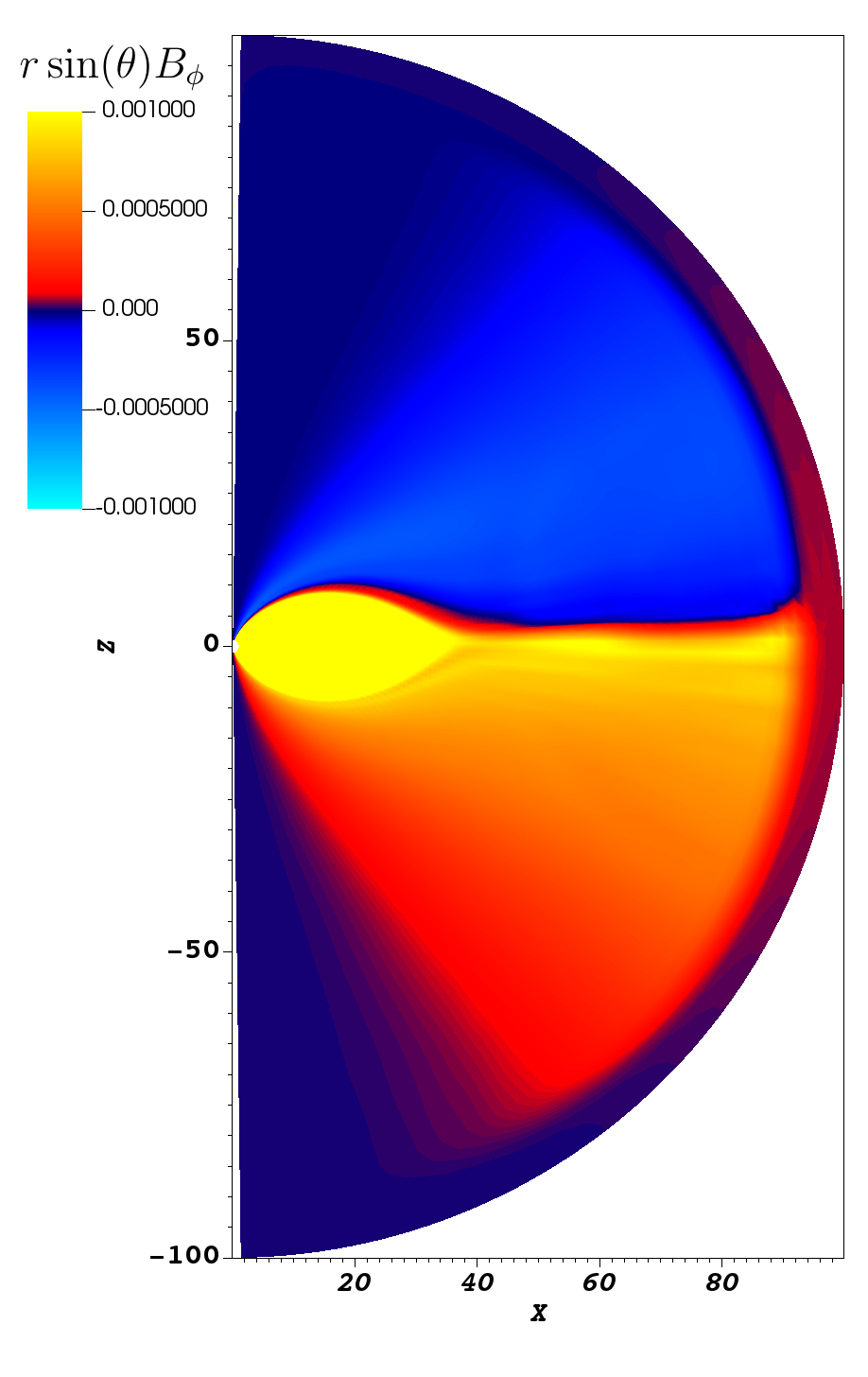}\\
\rotatebox{90}{$  \hspace{4em} C_{2}=0.1$}
\includegraphics[width=0.22\textwidth,height=0.22\textheight]{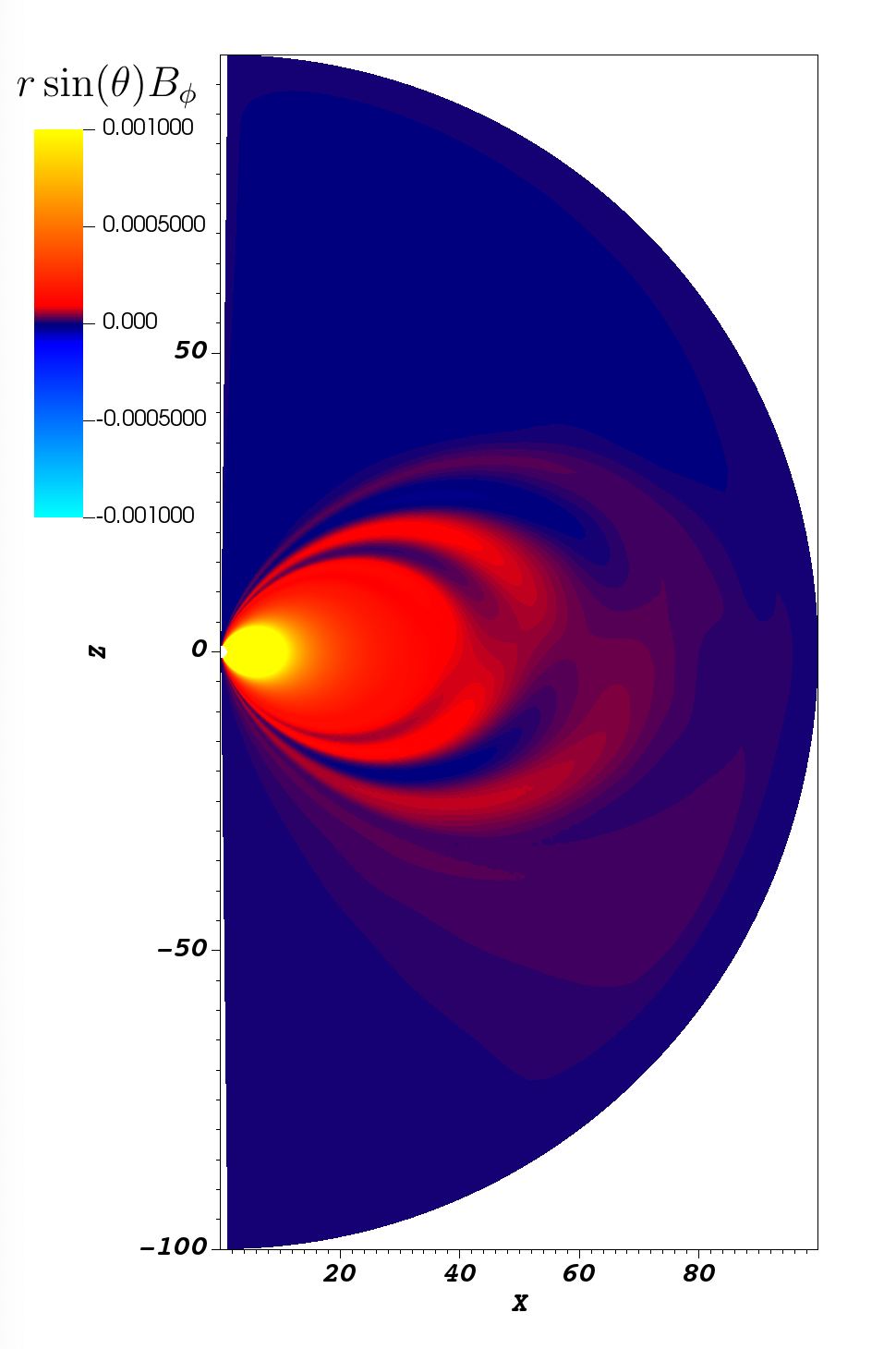}
\includegraphics[width=0.22\textwidth,height=0.22\textheight]{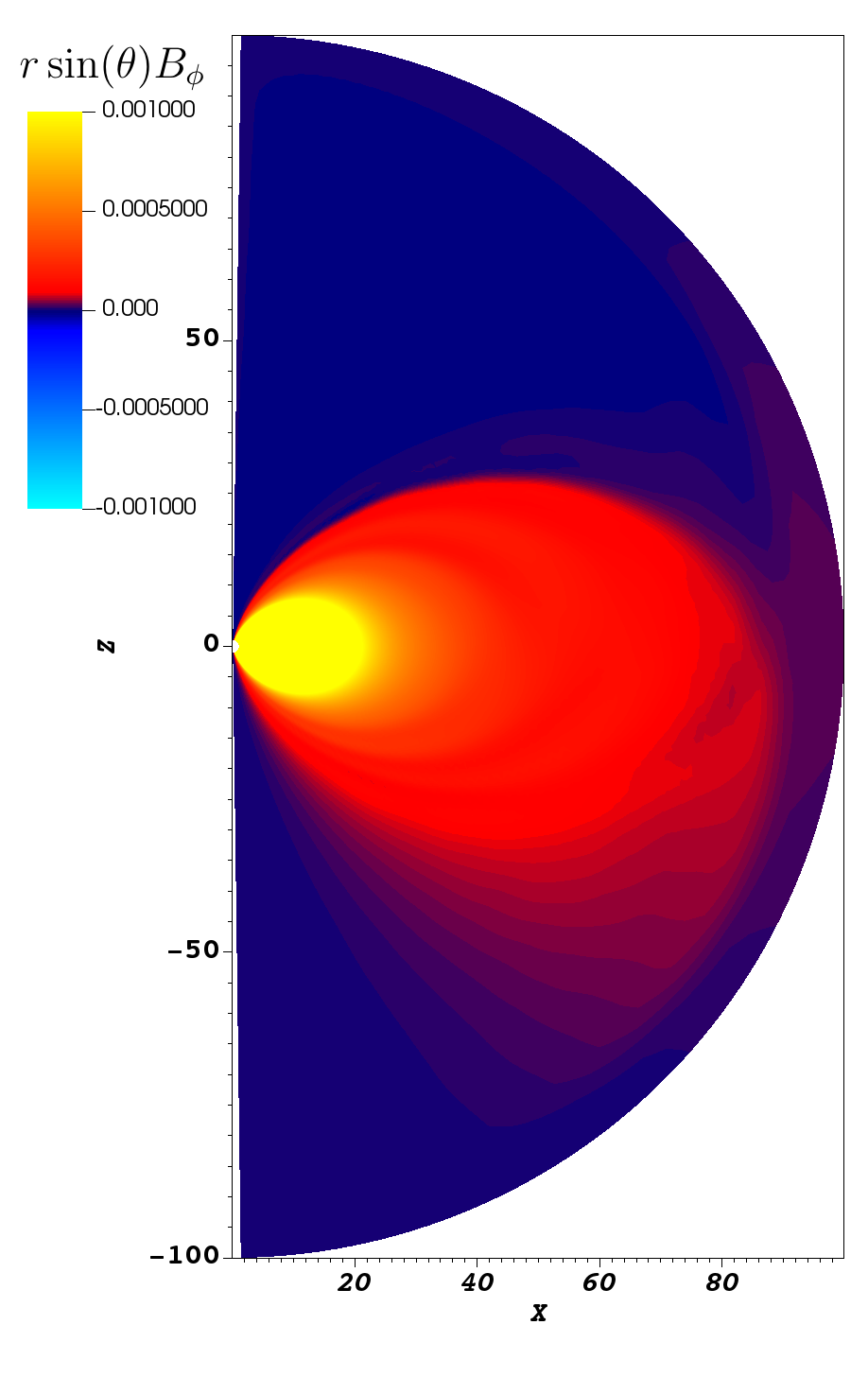}
\includegraphics[width=0.22\textwidth,height=0.22\textheight]{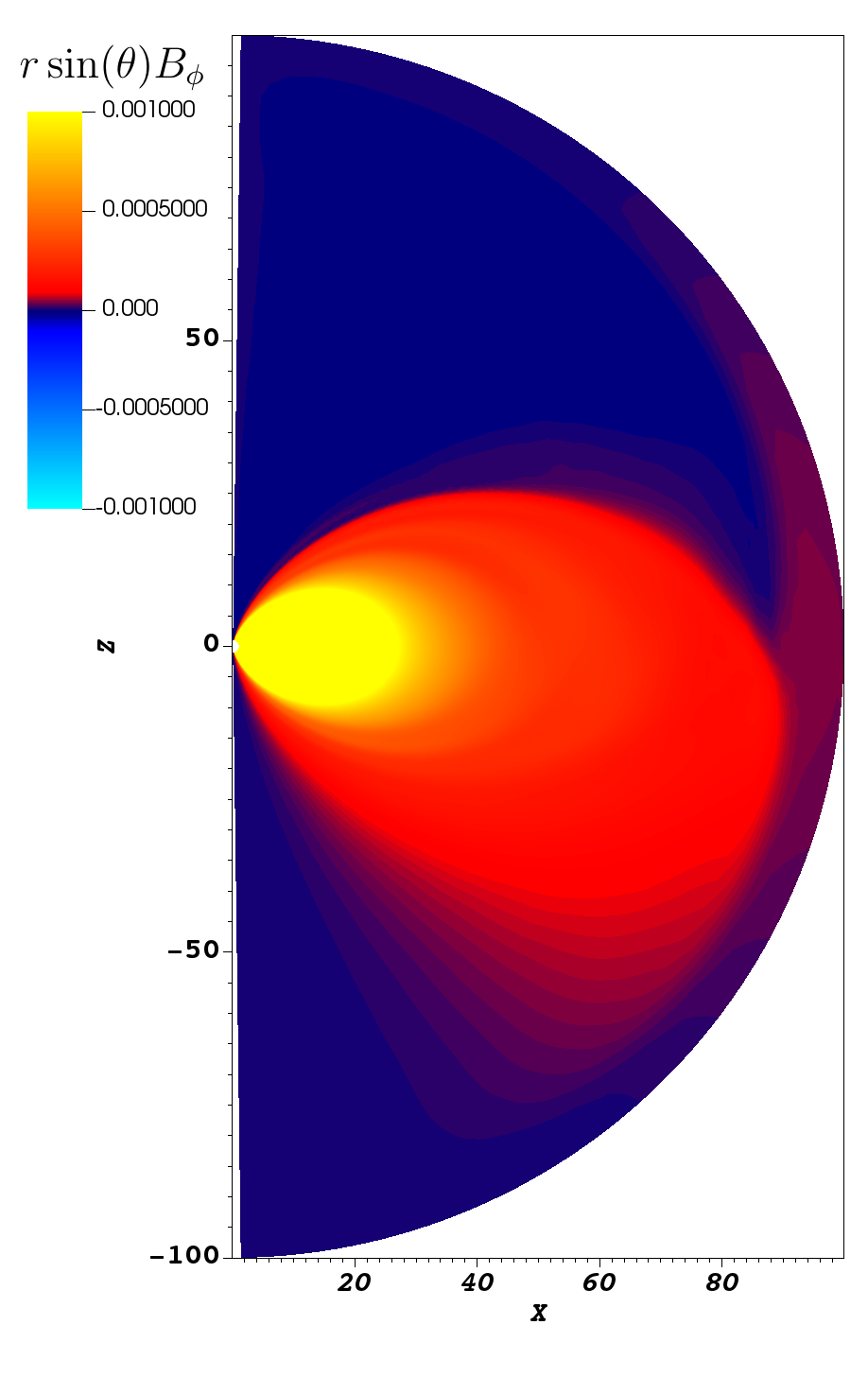}
\includegraphics[width=0.22\textwidth,height=0.22\textheight]{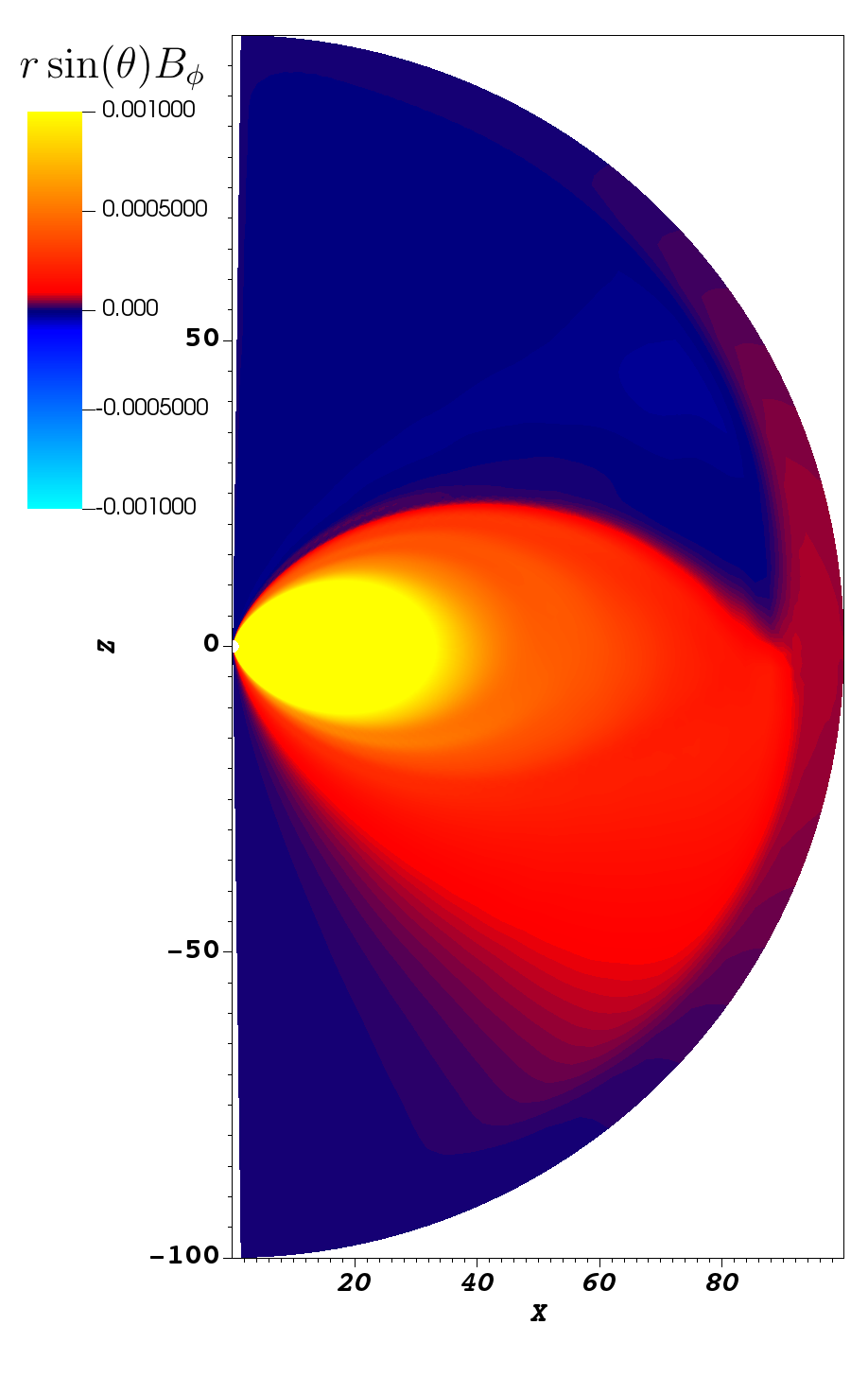}
\rotatebox[origin=c]{0}{$C_{1}=0.01$}\hspace*{6em}
\rotatebox[origin=c]{0}{$C_{1}=0.1$}\hspace*{6em}
\rotatebox[origin=c]{0}{$C_{1}=0.2$}\hspace*{6em}
\rotatebox[origin=c]{0}{$C_{1}=0.3$}
\caption{ A mosaic of  $r \sin(\theta)B_{\phi}$ for different twist rotation parameters $C_1$ and $C_2$.  $C_{2}=0.2$ (top panel), $C_{2}=0.15$ (middle panel), and $C_{2}=0.1$ (bottom panel). $C_{1} = 0.01, 0.1,0.2,0.3$ from left to right for both panels at $t= 30 $, measured in  light crossing time ${R_{NS}}{/c}$).  }
\label{3} 
\end{figure}

For small $C_2$ all twisted configuration do not have a \LC\ ( first two panels in Fig. \ref{00} and also bottom row in Fig. \ref{3}). When we increase the spin parameter $C_2$, the picture changed qualitatively, bottom panel in Fig. \ref{00} (Also see middle row in Fig. \ref{3}).

To further verify the observations made in previous paragraphs, we plot scaled toroidal magnetic field for various permutations of $C_{1},C_{2}$ in Fig. \ref{3}, 

From the figures it is clear that the presence of the light cylinder and wind depends strongly on $C_{2}$.  For slow rotation $C_2 =0.1$ (bottom row) there is no \LC. The \LC\ appears near $C_{2} \geq C_{2, crit}  \approx 0.15$.  (To guide an eye, formation of the wind requires \LC\ and the corresponding formation  of a  Y-point. Images with no Y-point imply no wind and no \LC.) A more precise critical value of $C_{2, crit} $ could not be determined:  at the transition the \LC\ appears at large distances, reaching the  absorbing layer next to the outer boundary).  For larger $C_2 \geq C_{2, crit} $ the \LC\ is located closer to the star. We restrict our simulations to $C_2 \leq 0.3$.  At this moment the \LC\ is located nearly at the star's surface, but the relaxation time becomes very long (if compared with small $C_2$ cases).

% The usual trend in location of  light cylinder is that at least for moderate values of $C_{2}$, $R_{LC}$, if present, usually increases as simulation progresses. We also observe that the light cylinder appears at $C_2 \ge 0.1$ and for a given $C_1$ , $R_{LC}$ decreases as $C_2$ increases.
 
%  We observe gradual shrinkage of  \LC, with in some cases light cylinders disappearing as the simulations further proceeds. We illustrate this non-stationary behavior in Figure \ref{4} for the case of $C_2=0.30,C_1=0.05$ and $C_2=0.29,C_1=0.10$
    
\section{Discussion} 

We discuss an analytical model of a differentially rotating \NSs\ with twisted \mss. We find a new type of   solutions without  {\LC}s.
 Such configuration could still  be a pulsar, a rotating \NS:  it will be more like magnetar producing periodically modulated  radio emission driven by  magnetic reconnection \citep{2002ApJ...580L..65L,2020arXiv200616029L},  not by the rotational energy.

The present model is  mostly mathematical.  It requires that the twisting motion of the foot-points be comparable to the spin; and  the spin shear  must be of the particular shape, related to twist.  
It is a bit surprising that rotating \mss\ can still  be self-similar, since there is a special distance, the \LC. Our results indicate, that there is a special set of parameters without \LC, thus allowing for self-similar solution.

We see the value of the model in  that,  first,  that it provides a clear analytical example of a new types of solutions of the pulsar equation: rotating yet non-spinning down  configurations. Secondly, 
 the model might  have implication for numerical modeling of twisted, sheared and rotating \mss\ of magnetars. Those types of models usually start with stationary \NS, and then twist, shear  and rotate it.  Necessarily, in those simulations the twisting, shearing  and the rotation rate must be similar, as the simulations  are  limited in their dynamic range: they  {\it have to} deal with twisting rates similar, just somewhat smaller, to the 
rotation rate.

\section{ACKNOWLEDGEMENTS}

This work had been supported by NASA grants 80NSSC17K0757 and 80NSSC20K0910, NSF grants 1903332 and 1908590.
We would like to thank  Yiannis Contopoulos,   Kyle Parfrey  and Alexander Philippov  for discussions. 

\section{Data availability}
The data underlying this article will be shared on reasonable request to the corresponding author.

\bibliographystyle{apj}
  \bibliography{/Users/maxim/Home/Research/BibTex}

\begin{thebibliography}{26}
\expandafter\ifx\csname natexlab\endcsname\relax\def\natexlab#1{#1}\fi

\bibitem[{{Beskin}(2009)}]{BeskinBook}
{Beskin}, V.~S. 2009, {MHD Flows in Compact Astrophysical Objects: Accretion,
  Winds and Jets}

\bibitem[{{Beskin} {et~al.}(1988){Beskin}, {Gurevich}, \&
  {Istomin}}]{1988Ap&SS.146..205B}
{Beskin}, V.~S., {Gurevich}, A.~V., \& {Istomin}, I.~N. 1988, \apss, 146, 205

\bibitem[{{Contopoulos} {et~al.}(1999){Contopoulos}, {Kazanas}, \&
  {Fendt}}]{1999ApJ...511..351C}
{Contopoulos}, I., {Kazanas}, D., \& {Fendt}, C. 1999, \apj, 511, 351

\bibitem[{{Goldreich} \& {Julian}(1969)}]{GJ}
{Goldreich}, P. \& {Julian}, W.~H. 1969, \apj, 157, 869

\bibitem[{{Goldreich} \& {Reisenegger}(1992)}]{RG}
{Goldreich}, P. \& {Reisenegger}, A. 1992, \apj, 395, 250

\bibitem[{{Gourgouliatos} {et~al.}(2015){Gourgouliatos}, {Kondi{\'c}},
  {Lyutikov}, \& {Hollerbach}}]{2015MNRAS.453L..93G}
{Gourgouliatos}, K.~N., {Kondi{\'c}}, T., {Lyutikov}, M., \& {Hollerbach}, R.
  2015, \mnras, 453, L93

\bibitem[{{Grad}(1967)}]{1967PhFl...10..137G}
{Grad}, H. 1967, Physics of Fluids, 10, 137

\bibitem[{{Gruzinov}(1999)}]{Gruzinov99}
{Gruzinov}, A. 1999, ArXiv Astrophysics e-prints

\bibitem[{{Gruzinov}(2005)}]{2005PhRvL..94b1101G}
---. 2005, \prl, 94, 021101

\bibitem[{{Komissarov}(2006)}]{2006MNRAS.367...19K}
{Komissarov}, S.~S. 2006, \mnras, 367, 19

\bibitem[{{Lynden-Bell} \& {Boily}(1994)}]{1994MNRAS.267..146L}
{Lynden-Bell}, D. \& {Boily}, C. 1994, \mnras, 267, 146

\bibitem[{{Lyutikov}(2002)}]{2002ApJ...580L..65L}
{Lyutikov}, M. 2002, \apjl, 580, L65

\bibitem[{{Lyutikov}(2006)}]{2006MNRAS.367.1594L}
---. 2006, \mnras, 367, 1594

\bibitem[{{Lyutikov}(2011)}]{2011PhRvD..83l4035L}
---. 2011, \prd, 83, 124035

\bibitem[{{Lyutikov}(2013)}]{2013arXiv1306.2264L}
---. 2013, arXiv e-prints, arXiv:1306.2264

\bibitem[{{Lyutikov}(2015)}]{2015MNRAS.447.1407L}
---. 2015, \mnras, 447, 1407

\bibitem[{{Lyutikov}(2020{\natexlab{a}})}]{2020arXiv200616029L}
---. 2020{\natexlab{a}}, arXiv e-prints, arXiv:2006.16029

\bibitem[{{Lyutikov}(2020{\natexlab{b}})}]{2020JPlPh..86b9010L}
---. 2020{\natexlab{b}}, Journal of Plasma Physics, 86, 905860210

\bibitem[{{Michel}(1973)}]{1973ApJ...180L.133M}
{Michel}, F.~C. 1973, \apjl, 180, L133

\bibitem[{{Parfrey} {et~al.}(2012){Parfrey}, {Beloborodov}, \&
  {Hui}}]{2012MNRAS.423.1416P}
{Parfrey}, K., {Beloborodov}, A.~M., \& {Hui}, L. 2012, \mnras, 423, 1416

\bibitem[{{Parfrey} {et~al.}(2013){Parfrey}, {Beloborodov}, \&
  {Hui}}]{2013ApJ...774...92P}
---. 2013, \apj, 774, 92

\bibitem[{{Scharlemann} \& {Wagoner}(1973)}]{1973ApJ...182..951S}
{Scharlemann}, E.~T. \& {Wagoner}, R.~V. 1973, \apj, 182, 951

\bibitem[{{Shafranov}(1966)}]{Shafranov}
{Shafranov}, V.~D. 1966, Reviews of Plasma Physics, 2, 103

\bibitem[{{Spitkovsky}(2006)}]{2006ApJ...648L..51S}
{Spitkovsky}, A. 2006, \apjl, 648, L51

\bibitem[{{Thompson} {et~al.}(2002){Thompson}, {Lyutikov}, \& {Kulkarni}}]{tlk}
{Thompson}, C., {Lyutikov}, M., \& {Kulkarni}, S.~R. 2002, \apj, 574, 332

\bibitem[{{Wood} {et~al.}(2014){Wood}, {Hollerbach}, \&
  {Lyutikov}}]{2014PhPl...21e2110W}
{Wood}, T.~S., {Hollerbach}, R., \& {Lyutikov}, M. 2014, Physics of Plasmas,
  21, 052110

\end{thebibliography}

\end{document}